\documentclass[useAMS,usenatbib]{mn2e}

\pdfoutput=1

\usepackage{epsfig,amsmath,amssymb}
\usepackage{natbib}
\usepackage{bm}
\usepackage{multirow}

\def \ebv {\ensuremath{E(B-V)}}
\def \kms {\ensuremath{{\rm km\,s}^{-1}}}
\def \xsizesingle {84mm}
\def \xsizedouble {168mm}

\let\oldsqrt\sqrt
\def\sqrt{\mathpalette\DHLhksqrt}
\def\DHLhksqrt#1#2{
\setbox0=\hbox{$#1\oldsqrt{#2\,}$}\dimen0=\ht0
\advance\dimen0-0.3\ht0
\setbox2=\hbox{\vrule height\ht0 depth -\dimen0}
{\box0\lower0.4pt\box2}}

\def \ariii {[Ar\,{\textsc {iii}}]}
\def \ariv {[Ar\,{\textsc {iv}}]}

\def \cliii {[Cl\,{\textsc {iii}}]}
\def \fei {Fe\,{\textsc {i}}}
\def \feii {[Fe\,{\textsc {ii}}]}
\def \feiii {[Fe\,{\textsc {iii}}]}
\def \fev {[Fe\,{\textsc {v}}]}
\def \fevii {[Fe\,{\textsc {vii}}]}
\def \hi {H\,{\textsc {i}}}
\def \hii {H\,{\textsc {ii}}}
\def \hei {He\,{\textsc {i}}}
\def \heii {He\,{\textsc {ii}}}

\def \ni {[N\,{\textsc {i}}]}
\def \nii {[N\,{\textsc {ii}}]}

\def \nai {Na\,{\textsc {i}}}
\def \neiii {[Ne\,{\textsc {iii}}]}

\def \oi {[O\,{\textsc {i}}]}
\def \oii {[O\,{\textsc {ii}}]}
\def \oiii {[O\,{\textsc {iii}}]}
\def \sii {[S\,{\textsc {ii}}]}
\def \siii {[S\,{\textsc {iii}}]}

\def \caii {Ca\,{\textsc {ii}}}

\begin{document}

\title[Analysis of emission-line galaxies using MFICA]{Classification
  and analysis of emission-line galaxies using mean field independent
  component analysis}

\author[J.~T.~Allen et al.]{James T.\ Allen,$^{1,2}$\thanks{E-mail:
    j.allen@physics.usyd.edu.au} Paul C.\ Hewett,$^1$ Chris
  T.\ Richardson,$^3$ Gary J.\ Ferland$^4$ and \newauthor Jack A.\
  Baldwin$^3$\\
  $^1$Institute of Astronomy, University of Cambridge, Madingley Road,
  Cambridge CB3 0HA\\
  $^2$Sydney Institute for Astronomy, University of
  Sydney, 44--70 Rosehill St, Redfern, NSW 2016, Australia\\
  $^3$Physics \& Astronomy Department, Michigan State University, East
  Lansing, MI 48824-1116, USA\\
  $^4$Physics \& Astronomy Department, University of Kentucky,
  Lexington, KY 40506-0055, USA}

\maketitle

\begin{abstract}
We present an analysis of the optical spectra of narrow emission-line
galaxies, based on mean field independent component analysis (MFICA),
a blind source separation technique.  Samples of galaxies were drawn
from the Sloan Digital Sky Survey (SDSS) and used to generate compact
sets of `continuum' and `emission-line' component spectra.  These
components can be linearly combined to reconstruct the observed
spectra of a wider sample of galaxies.  Only 10 components -- five
continuum and five emission line -- are required to produce accurate
reconstructions of essentially all narrow emission-line galaxies to a
very high degree of accuracy; the median absolute deviations of the
reconstructed emission-line fluxes, given the signal-to-noise ratio
(S/N) of the observed spectra, are 1.2--1.8$\sigma$ for the strong
lines.  After applying the MFICA components to a large sample of SDSS
galaxies we identify the regions of parameter space that correspond to
pure star formation and pure active galactic nucleus (AGN)
emission-line spectra, and produce high S/N reconstructions of these
spectra.

The physical properties of the pure star formation and pure AGN
spectra are investigated by means of a series of photoionization
models, exploiting the faint emission lines that can be measured in
the reconstructions.  We are able to recreate the emission line
strengths of the most extreme AGN case by assuming the central engine
illuminates a large number of individual clouds with radial distance
and density distributions, $f(r)\propto r^\gamma$ and $g(n)\propto
n^\beta$, respectively.  The best fit is obtained with $\gamma = -0.75$
and $\beta = -1.4$.  From the reconstructed star formation spectra we
are able to estimate the starburst ages.  These preliminary
investigations serve to demonstrate the success of the MFICA-based
technique in identifying distinct emission sources, and its potential
as a tool for the detailed analysis of the physical properties of
galaxies in large-scale surveys.
\end{abstract}


\section{Introduction}
 
The optical and ultraviolet emission-line spectra of galaxies have
proven to be a valuable source of information regarding the physical
conditions that prevail within such objects.  However, the majority of
diagnostics and analysis techniques currently in use focus on
measurements of a few small regions of each spectrum, discarding the
remaining information.  With the greatly increased quality and
quantity of data made available by recent extragalactic surveys, new
analysis techniques that make use of all available data are required
in order to obtain a more detailed picture of the physical properties
and evolution of galaxies.

The classification and analysis of active galactic nuclei (AGN) and
star formation (SF) in narrow emission-line galaxies is one area in
which updated analysis techniques have the potential to provide new
insights into the physical processes that govern these objects.  The
well-established correlations between the properties of supermassive
black holes (SMBHs) and those of their host galaxies
(e.g.\ \citealp{Magorrian98,FM00,KG01,HR04}) suggest a strong
evolutionary link between the two.  Feedback processes, in which
galaxy-scale winds propelled by the AGN heat and expel the
interstellar medium, shutting down both star formation and further
black hole accretion, are often invoked to explain this link
(e.g.\ \citealp{Granato04}; \citealp*{SDH05}; \citealp{Croton06}).
Direct tests of such models have proved challenging, although some
progress has been made in measuring the time delays between star
formation and AGN activity \citep{Schawinski07,Wild10}, allowing a
comparison to the feedback timescales predicted by the simulations.

In order to investigate the relationship between AGN and SF across
modern large-scale surveys, a necessary first step is to accurately
identify each process.  The most commonly used methods to distinguish
AGN from SF in optical spectra date back to the work of
\citet*[][hereafter BPT]{BPT81} proposing the use of a number of line
ratio diagrams, in particular \oiii\,$\lambda5008 / $H$\beta$
vs.\ \nii\,$\lambda6585 / $H$\alpha$, to establish the ionizing source
powering an observed emission-line spectrum.  Additional line ratio
diagrams were introduced by \citet{VO87}.

Subsequent advances in photoionization modelling, combined with the
large datasets provided by the Sloan Digital Sky Survey
(SDSS; \citealp{SDSS}), have allowed more quantitative BPT-based
classifications to be made.  In particular, diagnostic lines were
defined in the BPT plane by \citet{Kewley01b} and \citet{Stasinska06},
based on photoionization modelling, while \citet{Kauffmann03}
presented an empirical classification based on a large sample of
galaxies from the SDSS.  \citet{Kewley06} further extended these
schemes by defining the region between the \citet{Kewley01b} and
\citet{Kauffmann03} lines as the `composite' region, in which each
galaxy is expected to host both SF and an AGN.  Many related diagnostic
methods have been proposed that make use of a variety of line ratios,
in some cases combining these with other properties of the galaxies
(e.g.\ \citealp*{DPW85,OTV92}; \citealp{Lamareille04,Lamareille10};
\citealp*{MHL11}; \citealp{Yan11,Juneau11}).

Notwithstanding the undoubted success of the BPT-based diagnostic
methods in separating AGN-dominated from SF-dominated galaxies, fixed
boundaries and bi-modal classifications are normally involved.  In
practice, for a wide range of applications, we wish to quantify the
contribution from each source to each galaxy, over the full range from
100 per cent AGN to 100 per cent SF, including those cases where only
a weak contribution from one of the sources is present.  Such
measurements will prove invaluable in, for example, studies of the
feedback mechanisms that relate supermassive black hole properties to
those of their host galaxies, in which case a full census of SF and
AGN properties will allow a far more sensitive analysis of the
connection between these two sources.

In order to make possible these more sensitive measurements, we must
develop techniques that incorporate all the information contained
within each optical spectrum.  Additionally, information from a large
number of spectra within a survey can be combined and analysed as a
single unit.  Blind source separation (BSS) techniques process data in
this way, deriving sets of component spectra that can be combined with
varying weights to reconstruct each of the input spectra.  The most
familiar BSS technique applied to astronomical spectra is principal
component analysis (PCA), which has seen use for more than two decades
\citep{MPS90, Francis92, Yip04a}. While the PCA-derived component
spectra can be used to provide approximations to the object
spectra, the interpretation of the individual component
spectra themselves has only rarely proved illuminating. The more ambitious goal
is to generate component spectra that relate to the underlying
physical constituents within a galaxy which can be analysed using
standard techniques to determine the physical conditions contributing
to the individual components.

More recently, other BSS techniques have been applied to the analysis
of UV/optical spectra, including independent component analysis (ICA;
\citealp{Lu06}) and non-negative matrix factorisation (NMF;
\citealp{BR07,Allen11}).  Here we present an application of mean field
independent component analysis (MFICA), a BSS technique previously
unused in astronomy, to the analysis of SDSS spectra.  MFICA is
applied to a sample of narrow emission-line galaxies, generating a
small number of component spectra that, along with a corresponding set
of weights, can be used to reconstruct the spectrum of each object in
the sample.  This approach allows for the straightforward
identification of spectra corresponding to pure SF and pure AGN, as
well as the generation of high signal-to-noise ratio (S/N) examples of
such spectra, which in turn can be analysed to determine their
detailed physical properties.

The primary goal of this paper is to describe the MFICA technique and
how the MFICA spectral components can be used to trace
physically-significant trends, parametrized as loci, in the spectra of large samples
of emission-line galaxies. We discuss our results for the SDSS sample
primarily to illustrate the method and assess how well it works. In Section~\ref{s:mfica} we
introduce MFICA, and in Section~\ref{s:samples} we describe the galaxy
samples used.  Section~\ref{s:method} describes the methods used to
generate component spectra, fit these components to observed spectra,
and identify the regions of parameter space corresponding to SF and
AGN\@.  In Section~\ref{s:results} we present our results, including a
preliminary investigation of the physical properties of galaxies
within the SF and AGN regions.  We summarise our conclusions in
Section~\ref{s:conclusions}.  Vacuum wavelengths are used throughout
the paper.

\section{Mean field independent component analysis}

\label{s:mfica}

Blind source separation (BSS) techniques are used to rewrite a data
matrix, ${\bf V}$, as the product of a set of components, ${\bf S}$,
and weights, ${\bf A}$:
\begin{equation}
\label{e:bss}
{\bf V} = {\bf AS}.
\end{equation}

In the context of this work, ${\bf V}$ is an $n \times m$ array of
flux measurements for $n$ different galaxies at $m$ wavelengths, ${\bf
  S}$ is an $r \times m$ array of the $r$ component spectra over the
same wavelengths, and ${\bf A}$ is an $n \times r$ array of the
corresponding weights for each galaxy.  For any individual galaxy, the
observed spectrum is written as a linear combination of the $r$
components.  In the case that $r < n$, the equality in
equation~\ref{e:bss} is an approximation, and the product ${\bf AS}$
can be viewed as a reconstruction of the original data.  The choice of
$r$ depends on a number of factors, including the expected nature of
the components, the S/N of the input data, and the particular purpose
of the analysis.

Mean field independent component analysis (MFICA; \citealp{HS02,OW05})
is a BSS technique that produces a small number of components that can
be used to provide reconstructions of the individual input spectra.
MFICA imposes a prior on the components, $P\left({\bf S}\right)$, and
combines $P\left({\bf S}\right)$ with the error in the reconstructions
to maximize the likelihood of the parameters:
\begin{equation}
  \label{e:pvasigma}
  P\left({\bf V}|{\bf A},{\bm \Sigma}\right) = \int {\rm d}{\bf S}
  P\left({\bf V}|{\bf A},{\bm \Sigma},{\bf S}\right)
  P\left({\bf S}\right),
\end{equation}
where
\begin{equation}
  \label{e:pvasigmas}
  P\left({\bf V}|{\bf A},{\bm \Sigma},{\bf S}\right) = \left({\rm det}\,
  2\pi{\bm \Sigma}\right)^{-\frac{N}{2}}
  e^{-\frac{1}{2}{\rm Tr}\left({\bf V}-{\bf AS}\right)^T{\bm
      \Sigma}^{-1}\left({\bf V}-{\bf AS}\right)}, 
\end{equation}
where $N$ is the number of input spectra, and ${\bm \Sigma}$ is the
noise covariance.  ${\bm \Sigma}$ does not need to be specified in
advance, and is calculated from the data along with {\bf A} and {\bf
  S}.  The parameters {\bf A} and {\bf S} provide a full description
of the noise-free spectra, while ${\bm \Sigma}$ describes the noise in the
observations.  In this work, ${\bm \Sigma}$ is taken to be a scalar.
In essence, MFICA derives a combination of {\bf A}, {\bf S} and ${\bm
  \Sigma}$ that combine to explain the input data {\bf V}, while
preferentially selecting values for the individual pixels in {\bf S}
that maximize the chosen $P\left({\bf S}\right)$.  Note that unlike
many ICA techniques, MFICA does not address the issue of statistical
independence.

The prior $P\left({\bm S}\right)$, which can in principle take almost
any form, can be used to place constraints on the components.
Restrictions can also be placed on the mixing matrix, {\bf A}.  In
particular, {\bf A} and {\bf S} can both be constrained to be
non-negative; such a constraint is appealing in the context of
spectroscopic observations where the physical emission signatures are
expected to obey such a restriction naturally.

\section{Galaxy sample}
\label{s:samples}

SDSS DR7 \citep{SDSSDR7} galaxy samples at redshift $z \simeq 0.1$
provide large numbers of spectra, with moderate S/N, covering a
restframe wavelength interval that includes \oii$\,\lambda$3728 in the
blue, through to \ariii$\,\lambda$7137 in the red. A narrow redshift
interval at $z \simeq 0.1$ was thus chosen for the application of
MFICA to investigate the properties of narrow emission-line galaxies.

More specifically, a large sample of SDSS-classified galaxies was
selected with redshifts $0.10\leqslant z < 0.12$.\footnote{Objects in
  the range $0.111\leqslant z < 0.116$ were discarded, because of the
  coincidence of the strong 5578.5\,\AA\ sky-line with rest-frame
  \oiii\,$\lambda$5008 in the galaxies.} The redshift range is
sufficiently broad that a large sample of $\sim$$10^4$ galaxies is
available, while retaining a large common rest-frame wavelength range
for analysis. The spectra were also required to have at least 3800
`good' pixels, defined as those for which the SDSS noise array is
non-zero, and an $r$-band S/N of $15.0\leqslant {\tt SN\_R} < 30.0$.

\begin{figure}
\includegraphics[width=\xsizesingle]{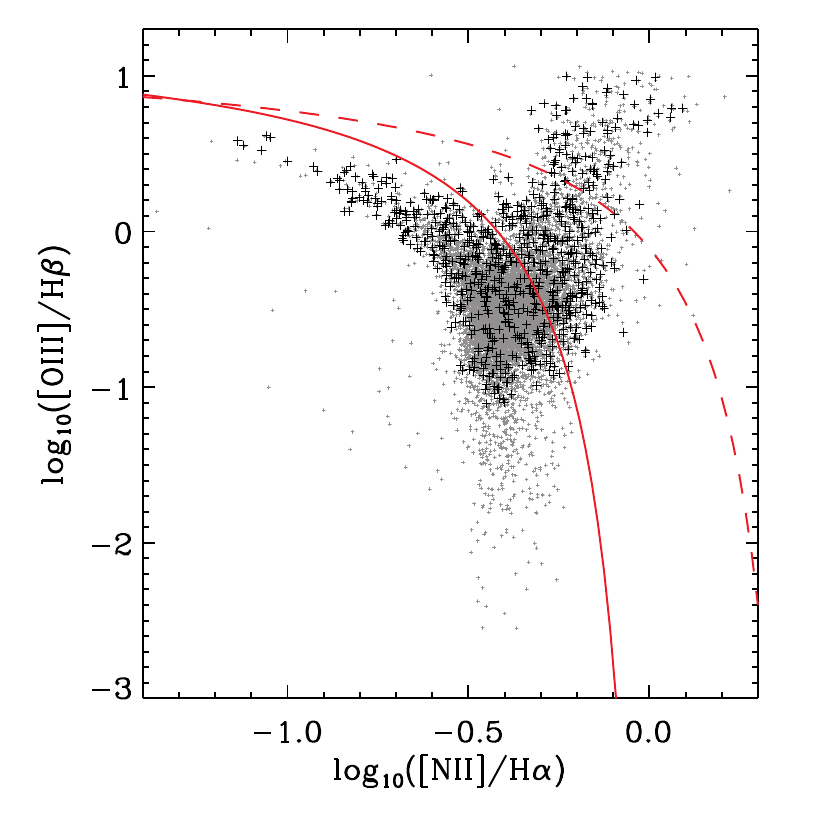}
 \caption{Positions of the input samples in the \oiii\,$\lambda5008 /
   $H$\beta$ vs.\ \nii\,$\lambda6585 / $H$\alpha$ BPT plane. Small
   grey points represent the complete emission-line galaxy sample,
   while large black points represent those galaxies used to generate
   the emission-line components. The solid and dashed red lines
   represent the classification curves defined in \citet{Kauffmann03}
   and \citet{Kewley01b}, respectively.}
 \label{f:sample_bpt}
\end{figure}

Emission-line galaxies were selected to have positive equivalent width
(EW), indicating emission, for each of H$\beta$, \oiii\,$\lambda$5008,
H$\alpha$ and \nii\,$\lambda$6585, with S/${\rm N} \geqslant 5.0$ for
the flux measurements of H$\beta$ and \nii\,$\lambda$6585.  The sample
was restricted to objects with an H$\alpha$ width $1.9\,{\text{\AA}}
\leqslant \sigma_{{\rm H}\alpha} < 3.5\,{\text{\AA}}$, and objects
with a visually-detected broad component to their H$\alpha$ emission
were removed. The resulting sample consists of galaxies with emission
lines detected at moderate S/N, and velocity widths in the interval
$\simeq 120-330$\,\kms, i.e. including $L^*$ and brighter
galaxies. Their positions in the \oiii\,$\lambda5008 / $H$\beta$
vs.\ \nii\,$\lambda6585 / $H$\alpha$ BPT plane are shown by the small
grey points in Fig.~\ref{f:sample_bpt}.

A sample of galaxies without emission lines was selected from those
with H$\alpha$ detected in absorption with S/${\rm N} \geqslant 3.0$
and no detected emission in \oii\,$\lambda$3728 or
\oiii\,$\lambda$5008.  A slightly broader redshift range,
$0.09\leqslant z < 0.13$, was used to increase the number of objects.

Before any further processing was carried out the spectra were
sky-subtracted using the algorithm presented by \citet{WH05b}, which
greatly improves the S/N at observed wavelengths $>7200$\,\AA.  All
spectra were corrected for Galactic dust reddening using the
\ebv\ measurements from \citet*{SFD98} and the Milky Way extinction
curve of \citet*{CCM89}.

The generation of components requires an accurate correction to
rest-frame wavelengths, and is sensitive to sub-pixel errors in this
correction.  For the emission-line galaxies, redshifts were remeasured
by fitting a set of three Gaussians to the
\nii\,$\lambda\lambda$6550,6585 and H$\alpha$ lines, after a
preliminary continuum subtraction.  As the primary focus of this work
is the emission line spectrum rather than the underlying continuum,
the H$\alpha$ redshift was adopted as the systemic redshift.  For the
galaxies without emission lines, the SDSS redshift measurements --
derived by cross-correlating the spectra with a series of templates --
were used.  The sky-subtracted spectra were shifted to their
respective rest frames, interpolating between the SDSS pixels to
ensure a precise rest-frame shift.  A common rest-frame wavelength
range of $3500\,{\text{\AA}} < \lambda < 8100\,{\text{\AA}}$ was
retained.

\subsection{Sample selection for component generation}

Subsets of the galaxy sample are required for the generation of
the MFICA-derived `continuum' and `emission-line' components. The
number of spectra required depends on the number of components
present in the object population and the S/N of the spectra.

The continuum components were derived using a modest-sized sample of
170 spectra.  This sample size is adequate because only a few such
components were being sought (Section~\ref{s:concomps}).  The galaxies
were selected to include an approximately even spread in continuum
colours between blue and red, i.e.\ between young (post-starburst) and
old (K-giant dominated) stellar populations.

The galaxy emission-line properties exhibit a greater variation,
requiring a somewhat larger sample of galaxies.  The first stage in
their selection was to place the objects in a classical
\oiii$/$H$\beta$ vs.\ \nii$/$H$\alpha$ BPT diagram. Galaxies
were then selected with probability inversely proportional to the
local density of galaxies in the BPT diagram, producing a sample
evenly populating the occupied portion of the diagram. Application of
somewhat tighter constraints on the spectrum S/N ($16.0\leqslant {\tt
  SN\_R} < 23.0$) and the H$\alpha$ emission-line width
($1.9\,{\text{\AA}} \leqslant \sigma_{{\rm H}\alpha} < 3.0$\,\AA)
resulted in a sample of 727 galaxy spectra. Their positions in the
\oiii$/$H$\beta$ vs.\ \nii$/$H$\alpha$
BPT plane are shown by the large black points in
Fig.~\ref{f:sample_bpt}. The original
sample consisted of 730 galaxies but, following visual inspection,
three galaxies with significantly dust reddened continua were
removed.  The presence of dust reduces the observed flux by a
wavelength-dependent factor; although the MFICA analysis is able to
account for moderate levels of dust, the most extreme objects can no
longer be described accurately by equation~\ref{e:bss} and so are
not included in the component generation.

All the spectra used in the component generation were normalised
according to their median flux, to prevent a small number of bright
galaxies from dominating the components.

\section{Method}
\label{s:method}

MFICA was used to generate components from subsets of the SDSS sample,
and these components were then fitted to the full sample of galaxies
in order to study their continuum and emission-line properties. The
analysis is described in detail in the following subsections; a brief
overview is given here. First, a set of five continuum components was
generated from a combination of galaxies with and without detected
emission lines (Section~\ref{s:concomps}). The components were
constructed in such a way as to avoid contamination by emission lines,
allowing them to be used to reconstruct and subtract the stellar
continuum in SDSS galaxy spectra. Subtracting the continuum in this
manner allows us to isolate each galaxy's emission-line spectrum with
minimal contamination from underlying stellar absorption
features. Recognising the importance of accurate continuum subtraction
in order to measure weak emission-line fluxes, a series of tests that
probe that accuracy of the MFICA-based continuum reconstructions are
described in Section~\ref{s:cont_accuracy}. The
continuum-subtracted spectra were then used to generate a set of five
MFICA components that can be combined together to describe the
emission line spectrum of any galaxy, including both AGN and SF
galaxies (Section~\ref{s:emcomps}). The individual components do not
represent distinct physical sources; rather, they are high S/N
representations of spectroscopic traits that appear in galaxy spectra,
and which include weaker emission lines that are now largely free of
contamination by the underlying galaxy.

The five continuum and five emission-line components were then fitted
to each individual galaxy spectrum in the full sample
(Section~\ref{s:fitting}). The results of this fitting allow us to
characterise the emission-line properties of each galaxy in a
five-dimensional space, independent of the continuum
properties. Finally, Section~\ref{s:loci} describes the identification
of two loci of galaxies running through this five-dimensional
space. One locus is identified with SF galaxies, and the other with
AGN. The position of a galaxy within one of these loci is determined
by its emission-line properties, and hence by the underlying physical
properties of that particular SF or AGN galaxy.

\subsection{Generating continuum components}
\label{s:concomps}

In principle, given enough galaxy spectra of extremely high S/N and
resolution, a BSS analysis should produce components that correspond
to stellar spectral types that make up loci of different ages and
metallicities in a Hertzsprung-Russell diagram. More realistically the
goal for spectra of the quality in the SDSS is the identification of
components that represent current star formation (O/B-star dominated),
intermediate age (post-starburst, A-star dominated) and old (K-giant
dominated), possibly with some additional age discrimination for the
intermediate age star-formation signatures.

Emission-line galaxies on their own are unsuitable for deriving
continuum components as the emission lines coincide with important
features in the underlying stellar spectrum.  This problem is
particularly pronounced for the Balmer series lines, which are seen as
strong absorption features in a range of main sequence stars.
However, galaxies with no emission lines also produce unsatisfactory
continuum components, as they do not include significant contributions
from young O/B stars.  In general, a sample that does not include a
significant contribution due to star formation of all ages will not
allow a BSS analysis to generate components that successfully
reconstruct the full range of star-formation ages.

\begin{figure*}
\includegraphics[width=\xsizedouble]{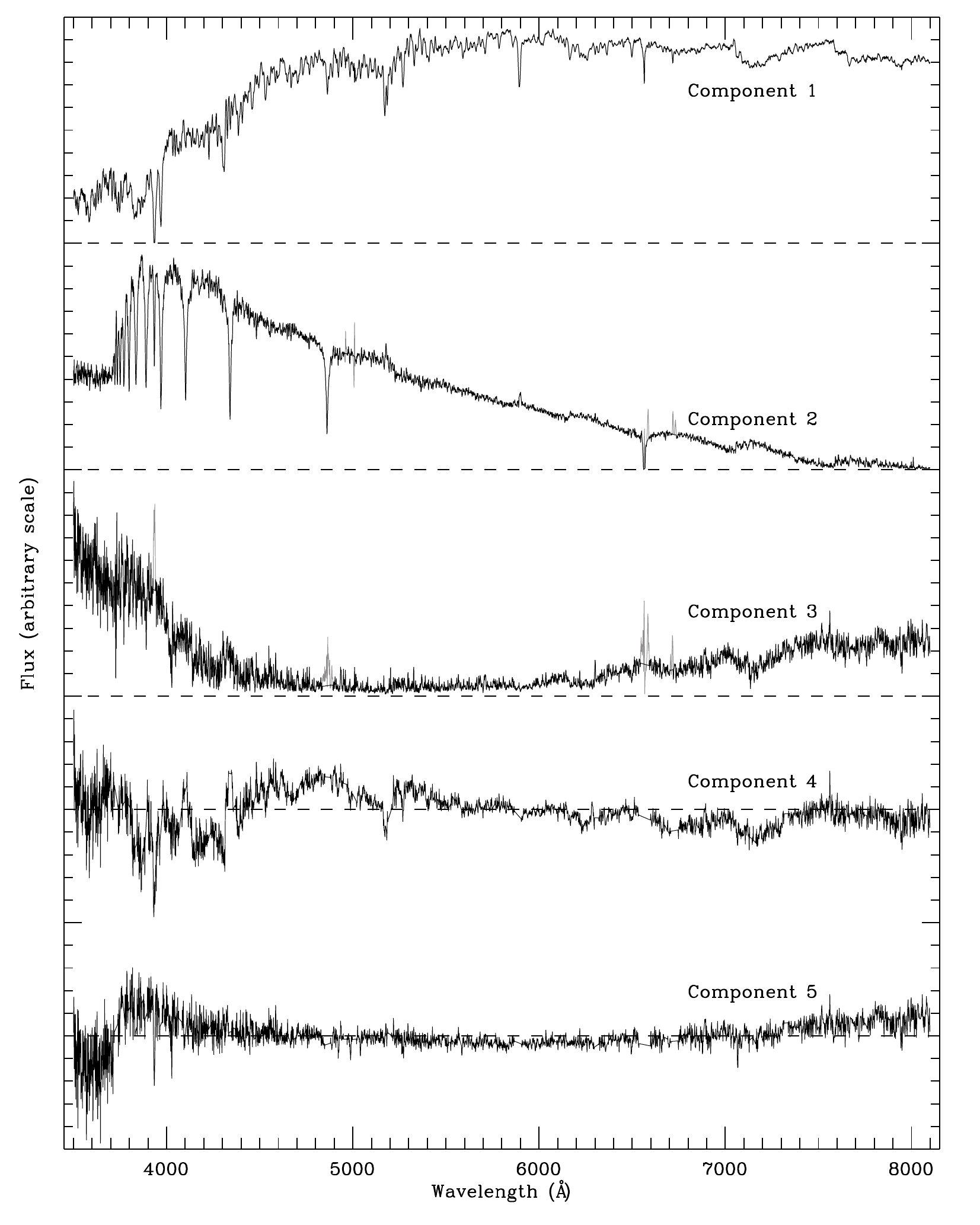}
 \caption{MFICA continuum components. 
   Components are offset for clarity; dashed lines mark successive
   zero points.  Components 1--3 were generated from a
   mixed sample of 170 galaxies without emission lines and 20 with
   emission lines; the final two `adjustment' components, numbers 4
   and 5, were
   generated from a mixed sample consisting of the same 170 galaxies
   without emission lines and 727 galaxies with emission lines. For
   components 1--3, the grey lines show the original
   components, while the black lines show the final versions after
   interpolating over a small number of narrow features.}
 \label{f:cont_comp}
\end{figure*}

Here, such limitations affecting the identification of continuum
components representing all star-formation ages were circumvented by
using a mixed sample of galaxies with and without emission lines.
First, a set of 20 emission-line galaxies, with particularly strong
contributions to their continua from young stars, was selected using a
preliminary MFICA analysis. A set of two MFICA components, which were
very similar in form to the top two components in
Fig.~\ref{f:cont_comp}, was generated from the 170 galaxies without
emission lines, and fitted to the 727 galaxies with emission
lines. The 20 selected were those with the highest fractional
contribution from the component corresponding to a young stellar
population. These 20 galaxies were added to the sample of 170 galaxies
without emission lines. The selection of galaxies with strong
young-star contributions allowed a significant signal from such stars
to be included in the sample with only a small number of emission-line
galaxies, ensuring the great majority of galaxies in the sample still
had no detected emission lines.

A set of seven components was
generated from the mixed sample, using the exponential prior:
\begin{equation}
\label{e:exponential_prior}
P\left(S\right) = \eta \exp \left(-\eta S\right) \Theta
\left(S\right),
\end{equation}
where $\Theta \left(S\right)$ is the Heaviside step function and
$\eta$ is a dimensionless parameter, taken in this case as $\eta=1$.
This prior was selected as it is zero for negative values of $S$, so
it produces non-negative component spectra, suitable for the
characterisation of emission sources.  Provided this condition is
satisfied, the resulting components do not depend strongly on the form
of the prior.  The seven components could be cleanly divided into
three that were dominated by the stellar continuum and four that were
dominated by the emission lines. The emission-line components were not
used in the following analysis -- the method described in
Section~\ref{s:emcomps} produced emission-line components that could
more easily be analysed, and at higher S/N -- but they played the role
of filtering out virtually all the emission-line signal, leaving the three
continuum components almost entirely free of contamination by emission
lines.

The number of components generated at this point was chosen to provide
physically meaningful continuum components with a clean separation
from the emission-line components.  Generating fewer components causes
some components to show a combination of continuum and emission-line
signatures, rather than being dominated by one or the
other. Generating a greater number of components from this sample
causes the continuum signal to be spread over more components in such
a way as to obscure the clear physical interpretation discussed below,
while providing only a minimal improvement in the accuracy of the
resulting reconstructions. The
choice of seven components was made after inspecting the results from
a range of numbers of components.

The performance of the MFICA in terms of the degree of cross-talk
between components is impressive. However, the presence of the very
high-contrast, narrow emission lines, combined with the limited S/N of
the galaxy spectra, means that some low-level contamination of the
continuum components by residual emission features is present.
Specifically, two of the three continuum-dominated components showed
some contamination, primarily at the location of the strongest
emission lines. The regions of contamination were identified by visual
inspection. A linear interpolation was then applied to produce
the final continuum components.  The parameters used in the
interpolation are listed in Table~\ref{t:cont_smooth}; for each
emission line the continuum between $\lambda_{\rm min}$ and
$\lambda_{\rm max}$ was set by interpolating between the median flux
levels in the $N_{\rm pix}$ pixels on either side of the region.

\begin{table}
  \centering
  \caption{Parameters used to interpolate over features in continuum
    components.}
  \label{t:cont_smooth}
  \begin{tabular}{ccccc}
    \hline
    Component & Feature & $\lambda_{\rm min}$ (\AA) &
    $\lambda_{\rm max}$ (\AA) & $N_{\rm pix}$ \\
    \hline
    \multirow{5}{*}{2} & \oiii\ & 4955.1 & 4967.6 & 15 \\
     & \oiii\ & 5000.9 & 5012.5 & 15 \\
     & H$\alpha$ & 6563.7 & 6568.3 & 1 \\
     & \nii\ & 6578.8 & 6589.5 & 1 \\
     & \sii\ & 6715.1 & 6736.7 & 15 \\
    \hline
    \multirow{4}{*}{3} & Ca K & 3927.8 & 3939.6 & 15 \\
     & H$\beta$ & 4836.7 & 4892.7 & 15 \\
     & H$\alpha$+\nii\ & 6544.1 & 6591.0 & 15 \\
     & \sii\ & 6704.2 & 6719.7 & 15 \\
    \hline
  \end{tabular}
\end{table}

The top three spectra in Fig.~\ref{f:cont_comp} show the original
components, as well as the results after interpolation. The three components
can be identified with old,
intermediate and young stellar populations (dominated by K, A and O
stars respectively), although the third of these components is
relatively noisy.  The O star component also shows an increase in flux
towards the red from a population of red supergiants, formed by the
rapid evolution of the most massive main sequence stars.

Although these three continuum components are very successful
in reconstructing the continua of observed galaxy spectra, some
systematic low-level residuals remain. There are two main effects that
cause these residuals. First, as a stellar population ages, its
spectral energy distribution (SED) approximates a blackbody with
successively lower temperatures, so its peak flux shifts to longer
wavelengths.  Each of the three MFICA-derived continuum components
necessarily represents an average over a range of ages, whereas the
SED of individual galaxies can be dominated by star formation of a
particular age. Both very young and old stellar populations produce
almost invariant signatures in spectra with the S/N and resolution of
the SDSS; the question is essentially how much of each component is
present. By contrast, the signature of intermediate-age,
post-starburst, populations produces significant changes with age as
the dominant stellar type evolves from late B-, through A- to early
F-type stars, and the three components cannot fully reflect these
changes on their own.  The second effect is that the relative
contributions of the three components to an individual galaxy is
largely driven by the overall shape of the SED but, for example, a red
SED can result from an old stellar population with little dust, or
from a younger population with a higher level of dust reddening.  The
spectra of these two possibilities will differ in features such as the
strength of the 4000-\AA\ break and the Balmer absorption lines and,
as the MFICA approach used here does not explicitly account for dust reddening,
these differences are not fully described by the existing three
components.

It would be possible to reduce the residuals to some extent by
increasing the number of components generated at the first step
described above. This approach is not followed here, for two
reasons. Firstly, the resulting components can no longer be clearly
identified with stellar populations of different ages, making
investigations of the sort described in Section~\ref{s:vespa} more
challenging. Secondly, a wider range of galaxies can be reconstructed
by generating additional components, to be used in combination with
the first three, from a sample that has a greater number of
emission-line galaxies; the inclusion of such galaxies in the input
sample ensures that the components are able to describe such
galaxies. We adopt this strategy here.

In order to reconstruct more accurately the spectra of galaxies
covering a wide range of stellar populations and levels of dust
reddening, an additional set of `adjustment' components was generated,
using the 170 galaxies without emission lines and all 727 with
emission lines.  The original three continuum components were fitted to the combined
sample of 897 galaxies, with the emission-line wavelength ranges
listed in Table~\ref{t:cont_mask} masked out, using the MFICA
algorithm with the components held fixed.\footnote{In practice the
  procedure produces results almost identical to performing a minimum
  $\chi^2$-fit of the three components to each galaxy spectrum.}
After subtracting the initial three-component continuum fit, the
masked `residual' spectra were used to generate further MFICA
components. When combined with the `mean' intermediate age
star-formation component, the new components allow the generation of
spectra corresponding to somewhat younger and older star
formation.

The positivity constraints on these additional components and their
weights were dropped, and a Laplace prior was used, defined as
\begin{equation}
\label{eq:laplace_prior}
P\left(S\right) = \frac{\eta}{2} \exp \left(-\eta |S|\right),
\end{equation}
again with $\eta=1$.  The Laplace prior allows both positive and
negative values, making it suitable for generating components that are
intended to describe deviations from a previous set of positive
components.  It was found that generating two additional components
was sufficient to fully reconstruct the observed continua, with any
further components producing no significant improvement.

\begin{table}
  \centering
    \caption{Wavelength ranges masked when generating continuum adjustment
      components.}
    \label{t:cont_mask}
    \begin{tabular}{ccc}
     \hline
     Emission line & $\lambda_{\rm min}$ (\AA) & $\lambda_{\rm max}$
     (\AA) \\
     \hline
     \oii & 3719.5 & 3737.5\\
     H$\kappa$ & 3748.1 & 3755.3\\
     H$\iota$ & 3765.6 & 3775.1\\
     H$\theta$ & 3794.8 & 3803.5\\
     H$\eta$ & 3831.3 & 3840.0\\
     \neiii & 3865.9 & 3873.4\\
     \hei+H$\zeta$ & 3883.5 & 3895.4\\
     \neiii+H$\epsilon$ & 3964.6 & 3975.5\\
     \sii & 4066.4 & 4081.9\\
     H$\delta$ & 4098.5 & 4108.3\\
     \fev & 4225.5 & 4235.5\\
     H$\gamma$ & 4335.1 & 4347.8\\
     \oiii & 4358.8 & 4369.1\\
     \hei & 4469.0 & 4476.5\\
     \feiii & 4655.5 & 4663.2\\
     \heii & 4683.1 & 4690.9\\
     \ariv & 4708.8 & 4716.6\\
     \ariv & 4737.5 & 4745.5\\
     H$\beta$ & 4850.1 & 4879.0\\
     \oiii & 4949.9 & 4967.7\\
     \oiii & 4997.7 & 5019.2\\
     \ariii+\ni & 5188.9 & 5207.0\\
     \cliii & 5514.6 & 5523.8\\
     \cliii & 5534.6 & 5543.9\\
     \oi & 5574.2 & 5583.5\\
     \nii & 5751.4 & 5761.0\\
     \hei+Na\,D & 5869.3 & 5904.5\\
     \fevii & 6082.9 & 6093.1\\
     \oi & 6296.8 & 6308.3\\
     \siii & 6308.6 & 6319.1\\
     \oi & 6360.2 & 6370.8\\
     H$\alpha$+\nii & 6537.0 & 6598.2\\
     \hei & 6673.4 & 6686.6\\
     \sii & 6708.2 & 6743.8\\
     \ariii & 7130.8 & 7143.7\\
     \feii & 7150.2 & 7164.1\\
     \oii & 7315.9 & 7337.8\\
     \ariii & 7746.7 & 7759.7\\
     \hline
    \end{tabular}
\end{table}

The regions that had been masked out when the adjustment components
were generated were interpolated over, using a simple linear
interpolation, to produce the final
components.  The two `adjustment' components are included in
Fig.~\ref{f:cont_comp}. Having removed the non-negativity constraint,
these two components do not have clear physical
interpretations. However, their presence does not negate the physical
interpretations of the first three components, which resulted from the
enforcement of non-negativity at that stage.

The continuum components were normalised such that the sum of the
squares of their pixel values is equal to unity. As noted above, dust
reddening is not explicitly accounted for in the
reconstructions, but in practice the range of levels of reddening in
the input spectra allows a similar range to be reconstructed
successfully by the MFICA components.  This point is discussed further
in Section~\ref{s:vespa}.

\subsection{Accuracy of the continuum components}

\label{s:cont_accuracy}

\begin{figure*}
\includegraphics[width=\xsizedouble]{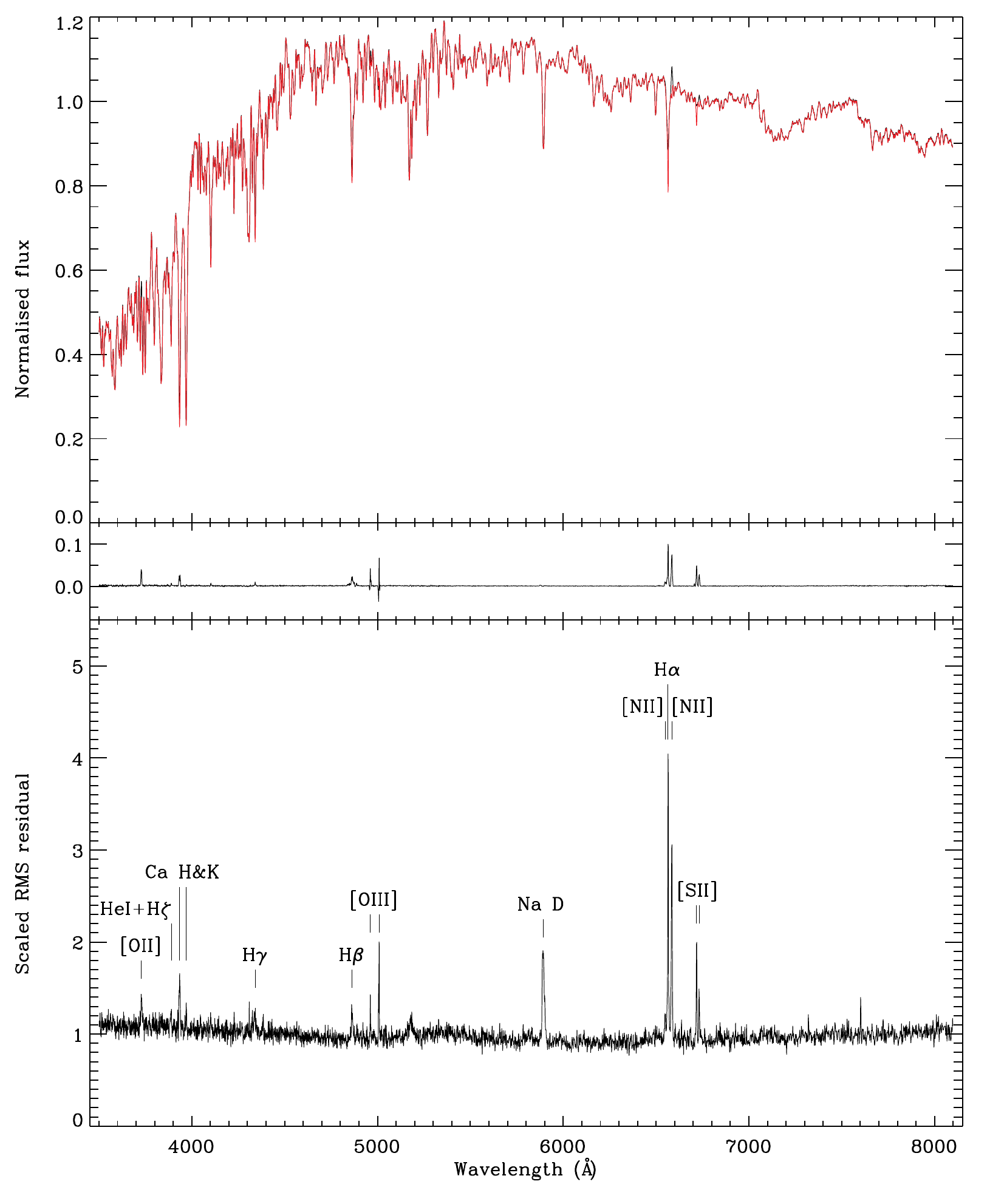}
 \caption{Top: mean normalised spectrum (black) of the galaxies
   without emission lines, and the corresponding mean MFICA
   reconstruction (red).  Middle: mean residuals after subtracting the
   MFICA reconstruction.  Bottom: RMS of the residuals normalised by
   the SDSS noise arrays. Features corresponding to known emission and
 absorption lines are labelled. Pixels with zero noise reported in the
 SDSS spectrum were removed from the RMS calculation, as was a strong
 artefact covering a further four pixels in one of the spectra.}
 \label{f:cont_fit}
\end{figure*}

The effectiveness of the MFICA technique in reconstructing the
underlying galaxy continua and photospheric absorption is evident from
consideration of the features present in the mean and root mean square
(RMS) of the galaxy minus reconstructions, for the 170 galaxies
without detected emission lines (Fig.~\ref{f:cont_fit}).  The RMS is
calculated after scaling the residuals by the SDSS noise arrays. The mean
residual, shown in the middle panel of Fig.~\ref{f:cont_fit}, appears
to show an essentially featureless continuum with no detectable
absorption features present. Weak emission lines, including \nii,
H$\alpha$, \oiii\ and H$\beta$, do however, appear to be present. The
positions of these features are marked in the RMS spectrum in the
bottom panel of Fig.~\ref{f:cont_fit}.

\begin{figure}
  \includegraphics[width=\xsizesingle]{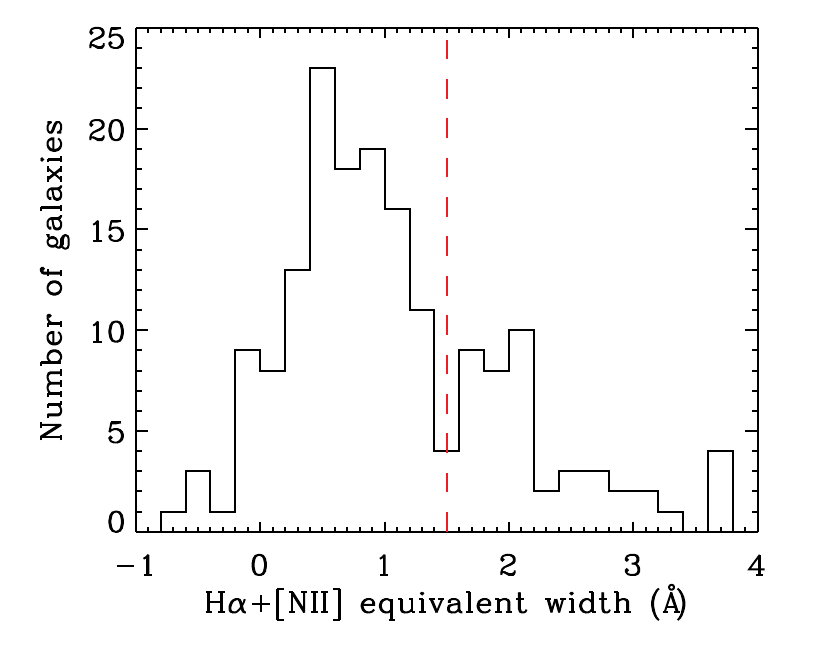}
  \caption{Histogram of the EW of the residuals in the H$\alpha$+\nii\
    spectral region, for the 170 galaxies without detected emission
    lines. Objects with ${\rm EW}>1.5$\,\AA, marked by the red dashed
    line, were included in the mean residual spectrum shown in
    Fig.~\ref{f:residual_liner}.}
  \label{f:residual_hist}
\end{figure}

\begin{figure*}
\includegraphics[width=\xsizedouble]{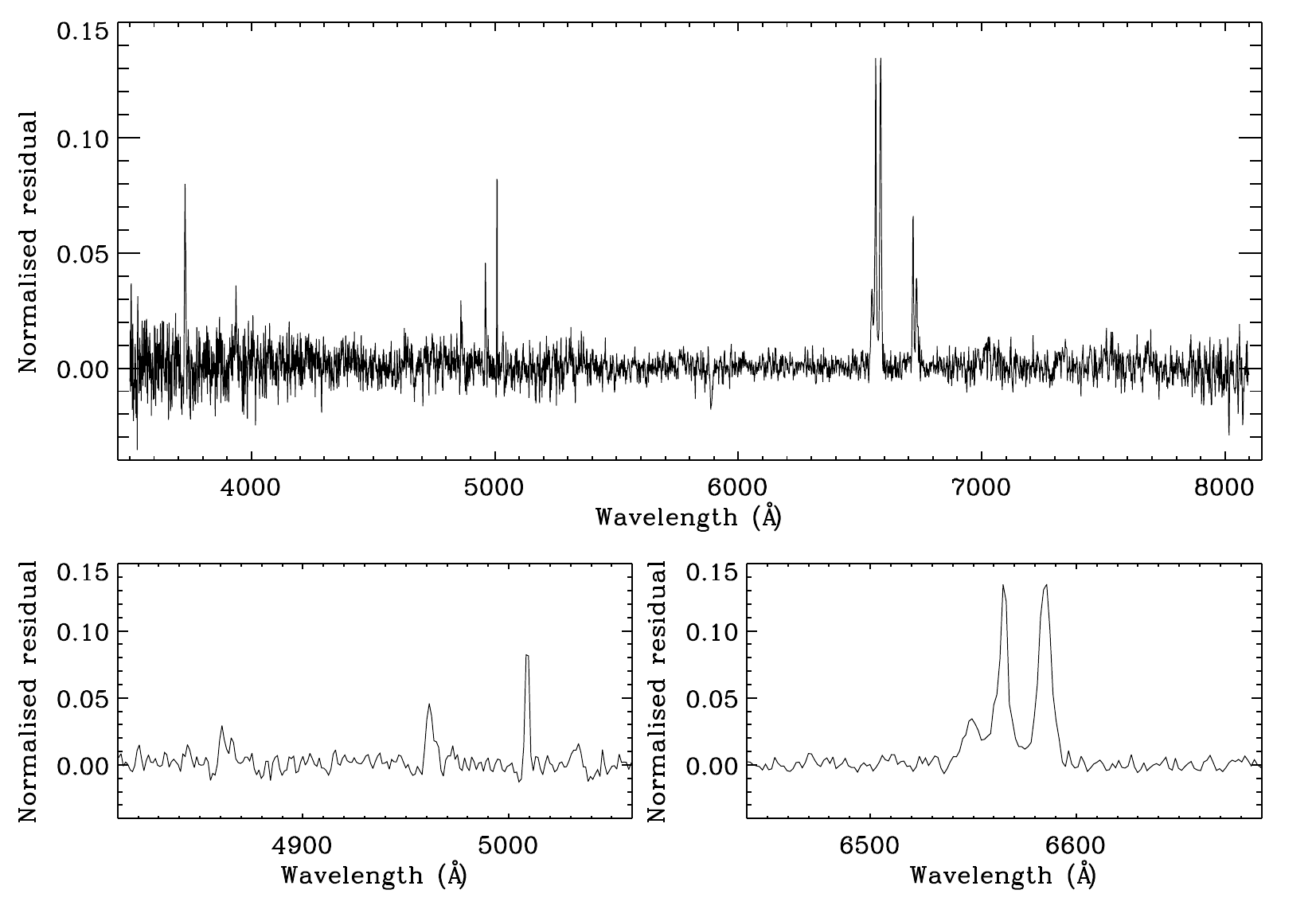}
 \caption{Mean residual spectrum, normalised to the median continuum
   level, of the 26 per cent of galaxies with the strongest residuals
 around H$\alpha$, from the sample without identified emission
 lines. The lower panels show the regions around H$\beta$ (left) and
 H$\alpha$ (right) in more detail, showing LINER-like emission-line
 ratios.}
 \label{f:residual_liner}
\end{figure*}

 The emission-line signature is, at first sight, unexpected given the
input galaxy sample was deliberately chosen to be free of galaxies
with detectable emission. Insight into the origin of the weak emission
line residuals comes from the distribution of the EW of the residuals
at the wavelengths of emission lines, shown in Fig.~\ref{f:residual_hist}. The distribution
consists of $\simeq$74 per cent of the galaxies centred on a residual
H$\alpha$+\nii\ EW of 0.8\,\AA\ ($\simeq
1.5\times10^{-16}$\,erg\,s$^{-1}$\,cm$^{-2}$), with a tail of
$\simeq$26 per cent of objects possessing larger positive flux
residuals. The flux residuals in the tail are well in excess of the
noise and a composite of the 26 per cent of galaxies with the largest
residuals shows a weak, but high S/N, emission-line spectrum
consistent with LINER flux ratios, with log(\nii/H$\alpha$) = 0.13 and
log(\oiii/H$\beta$) = 0.17. This composite is shown in
Fig.~\ref{f:residual_liner}. For comparison, we note that
\citet{Sarzi06} found weak emission lines with H$\beta$ EW in the
range 0.1--1.0\,\AA\ and line ratios consistent with LINERs to be
common in integral field observations of early-type galaxies.

The clear emission line signature in the mean residual spectrum thus
derives from the presence of a fraction of early-type galaxies that
possess weak emission line signatures that elude the SDSS
emission-line detection algorithm. Throughout the remainder of the
wavelength range, away from the emission lines, the mean residual is,
as expected, very small, not exceeding 0.5 per cent of the continuum
level. The presence of the sub-sample of emission-line galaxies is
also responsible for the excess RMS at the emission-line
wavelengths visible in the galaxy-minus-reconstruction RMS plot.

The RMS spectrum also displays a large increase in amplitude
coincident with the Na\,D\,$\lambda\lambda$5869.3,5904.5 doublet and a
less extreme increase coincident with the
\caii\,H\&K\,$\lambda\lambda$3969.6,3934.8 doublet. The distribution
of residuals
for the individual galaxies in the sample (at the wavelengths of at
Na\,D and \caii) is centred on zero, and shows the majority to possess
no significant residuals. Five of the 170 galaxies (3 per cent) show
much stronger absorption than predicted by the continuum
reconstructions. The absorption signature corresponds to the strong
resonant transitions from species that dominate the optical spectrum
of cool atomic gas in the interstellar medium of galaxies
\citep{Draine11}. Many of the genuine early-type galaxies in the
`continuum' galaxy sample do not possess cool atomic interstellar
medium, but the later-type galaxies (S0s and early-type spirals
without current star-formation) do possess interstellar gas that
manifests itself via absorption.  The limited number of continuum
MFICA components are unable to reproduce the additional absorption,
the observational signature of which is just a small number of
discrete absorption (negative) features in a small minority of the
input galaxies. It would be possible in principle to generate an
`absorption' component using the MFICA by increasing the number of
components for which the non-negativity constraint is not used, and
taking care not to mask absorption lines along with the emission
lines. However, given the finite S/N of the individual spectra and
very limited impact on the emission-line properties we have not
pursued such a course.

The other feature due to absorption lines that appears in the bottom
panel of Fig.~\ref{f:cont_fit} as having an unusually high RMS
residual is the Mg\,b band at 5180\,\AA. This feature is often used as
an alpha-element abundance tracer. It is seen to be strong in one of the
continuum adjustment components (component 4) shown in
Fig.~\ref{f:cont_comp}, suggesting that its strength has a large
scatter in our sample. The increased RMS indicates that the continuum
components do not fully account for this scatter. As for the Na\,D and
\caii\,H\&K features discussed above, the RMS residual at the
wavelength of Mg\,b could in principle be decreased by increasing the number of
components used. Again, we have not done so due to the very limited
benefit that would be gained.

We note the absence of increased RMS in the
wings of absorption features. The RMS peaks for the interstellar
features discussed above are no broader than the mean absorption
features themselves, while photospheric absorption features such as
\fei\,$\lambda$5270 are not visible in the RMS spectrum at all. As mentioned in
Section~\ref{s:samples}, the definition of the galaxy sample via a
narrow range in spectrum S/N, within a narrow redshift interval,
means that the galaxies possess a restricted range of luminosity and
hence mass. As a consequence, the variation in galaxy velocity
dispersion is small and produces no detectable effect on the form and
effectiveness of the MFICA continuum components when reproducing
photospheric absorption features. Had the input galaxy sample
contained a wider range of velocity dispersions, the limitations of
the MFICA approach in accounting for different widths of features
would have been expected to manifest itself in an increased RMS in the
wings of the absorption features.\footnote{If the
  continuum component definition is undertaken using a galaxy sample
  with a significant range of velocity-dispersion, or the continuum
  components are to be applied to galaxies with a significant range of
  velocity dispersion, galaxy spectra of continuum-components can be
  pre-smoothed as per the scheme for the emission lines described in
  Section~\ref{s:emcomps}.}

\begin{figure*}
\includegraphics[width=\xsizedouble]{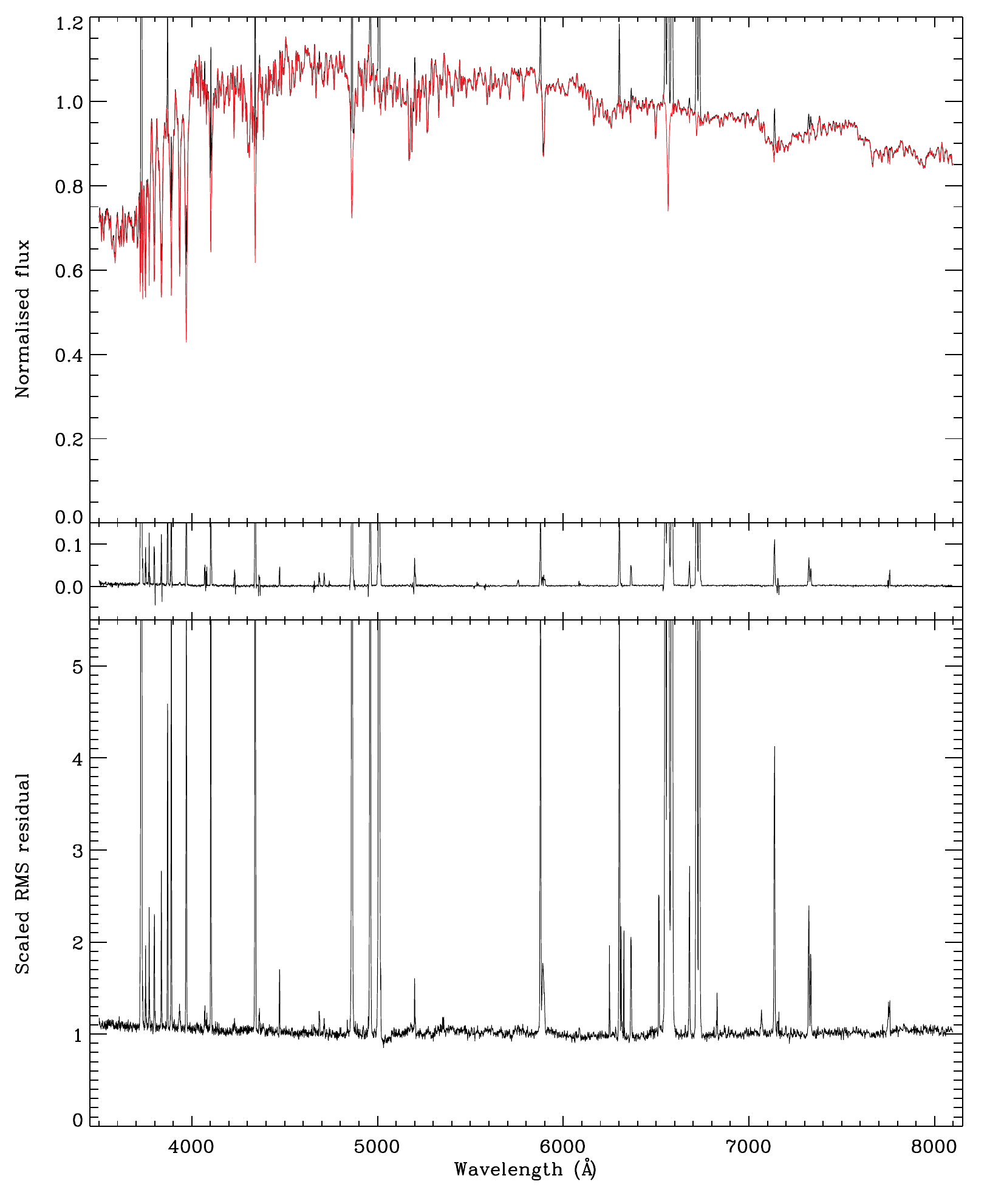}
 \caption{As Fig.~\ref{f:cont_fit}, but for the 727 emission-line
   galaxies used to generate the emission-line components.}
 \label{f:cont_fit_el}
\end{figure*}

The accuracy of the continuum reconstructions in emission-line
galaxies is illustrated in Fig.~\ref{f:cont_fit_el}, which shows the
mean continuum fit along with the mean residual and the RMS of the
residuals normalised by the noise. The sample of 727 galaxies used to
produce the emission-line components (Section~\ref{s:emcomps}) was
used to generate these
composites. As expected there are very strong
residuals at the positions of known emission lines, but in the
continuum regions the RMS residuals are
similar to those for the galaxies without identified emission lines,
indicating the continuum reconstructions are as accurate as can be
expected given the observational noise.

At the emission-line positions listed in Table~\ref{t:cont_mask} the
accuracy of the continuum components will be impacted by the masking
and interpolation used in their generation. To test the subsequent
effect on the emission-line components, the continuum components were
recalculated 150 times, following the procedure described in
Section~\ref{s:concomps}.  For each of the 150 repeats, an additional
masked region 10-\AA\ wide, chosen at random from the regions not
already masked, was added to the list in Table~\ref{t:cont_mask}. The
resulting continuum components were fitted to and subtracted from the
sample of 727 emission-line galaxies, which were then used to generate
new emission-line components following the procedure described in Section~\ref{s:emcomps}. For
simplicity, the median filtering to remove the low-level positive
offset described in Section~\ref{s:emcomps} was not performed; doing
so has a negligible effect on the emission line measurements. In order
to measure the potential effect of the masking procedure on an
emission line, four model spectra were generated from each of the 150
sets of emission-line components by combining them with weights
corresponding to each end of the star-forming and AGN loci described
in Section~\ref{s:loci}, i.e.\ the points s0, s2, a0 and a2 defined
in that Section. `Median components' were calculated by
taking the median value across the 150 realisations of each component
at each pixel. These median components were also combined using the
same sets of weights to produce
`median models', which were used as a baseline against which to
measure the effect of the additional masks. The change in the flux within the extra
10-\AA\ mask, relative to the median models, was measured and
normalised by the H$\beta$ flux; this change in flux was taken as a
measure of the effect of the mask on the flux of a coincident emission line.

\begin{figure*}
\includegraphics[width=\xsizedouble]{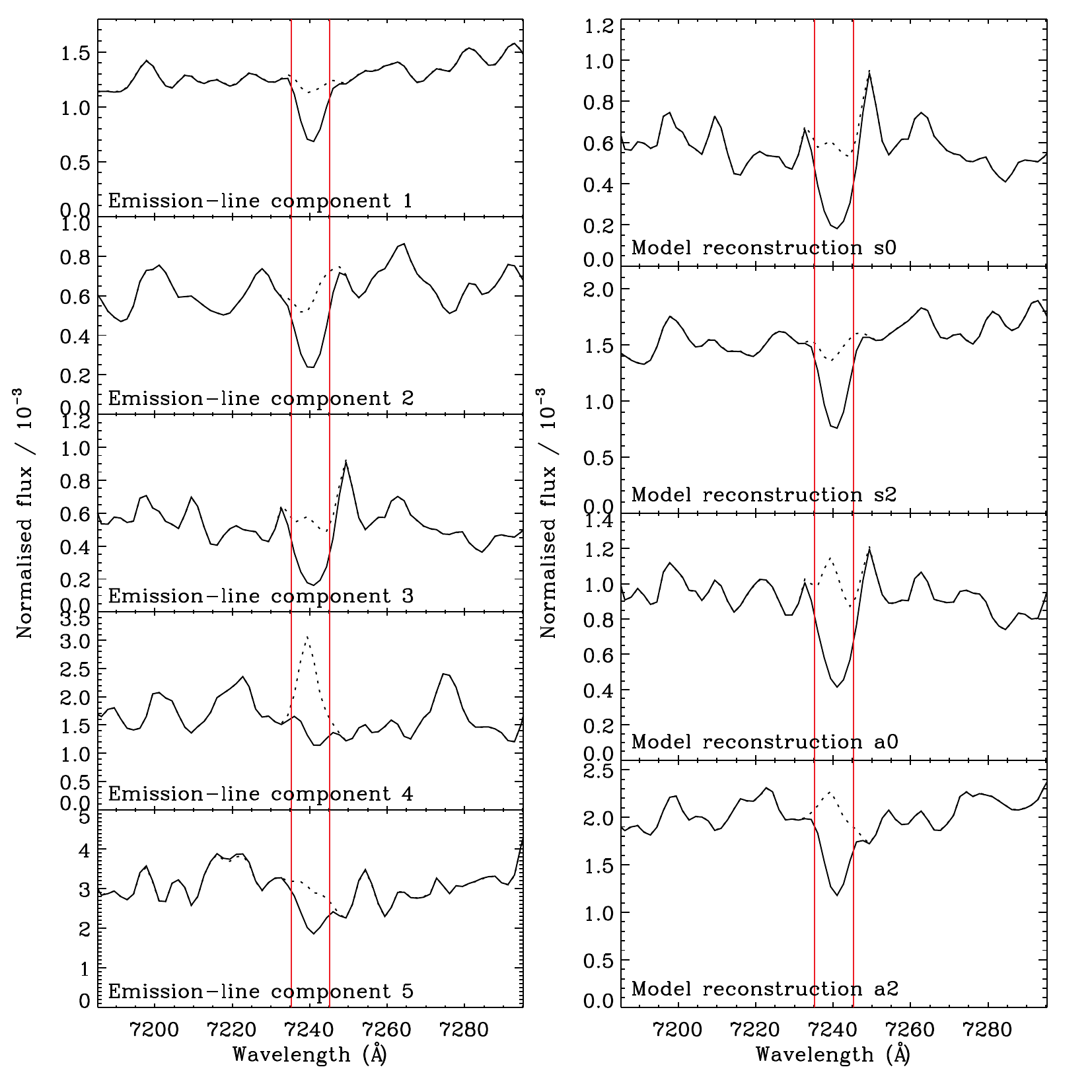}
 \caption{Left: Emission-line components (solid lines) generated after
an extra mask covering 7235.2--7245.2\,\AA\ was included in the
continuum component generation. The masked region is shown by vertical
red lines. The dotted lines show the median components over all 150
repeats. Right: Model spectra (solid lines) formed from the
emission-line components shown in the left panels. The median models
are shown as dotted lines. The difference between the solid and dotted
lines is taken as a measure of the typical error in emission line flux
caused by the masking and interpolation scheme. All components and
spectra are normalised relative to their peak flux. No genuine
emission lines are seen in the narrow wavelength range plotted.}
 \label{f:mask_test}
\end{figure*}

An example set of emission line components and model spectra is shown
in Fig.~\ref{f:mask_test}. In the example shown, the region between
7235.2 and 7245.2\,\AA\ was masked out when the fourth and fifth
continuum components were generated. The resulting emission-line components,
shown in the left panels of Fig.~\ref{f:mask_test}, are then seen to
deviate from the median values in this region. As a result, the
model spectra in the right panels also deviate from the median
levels. However, the error introduced by this mask is very small
relative to the peak flux in the components and models. Combining the
$4\times150=600$ flux measurements from all runs, the mean change in
flux relative to H$\beta$ is $2.0\times10^{-3}$, with a standard
deviation of $6.3\times10^{-3}$. Hence the error in the measured
emission line fluxes, introduced by the masking and interpolation carried
out during the continuum component generation, is constrained to be
less than 1 per cent of the H$\beta$ flux.

\begin{figure}
\includegraphics[width=\xsizesingle]{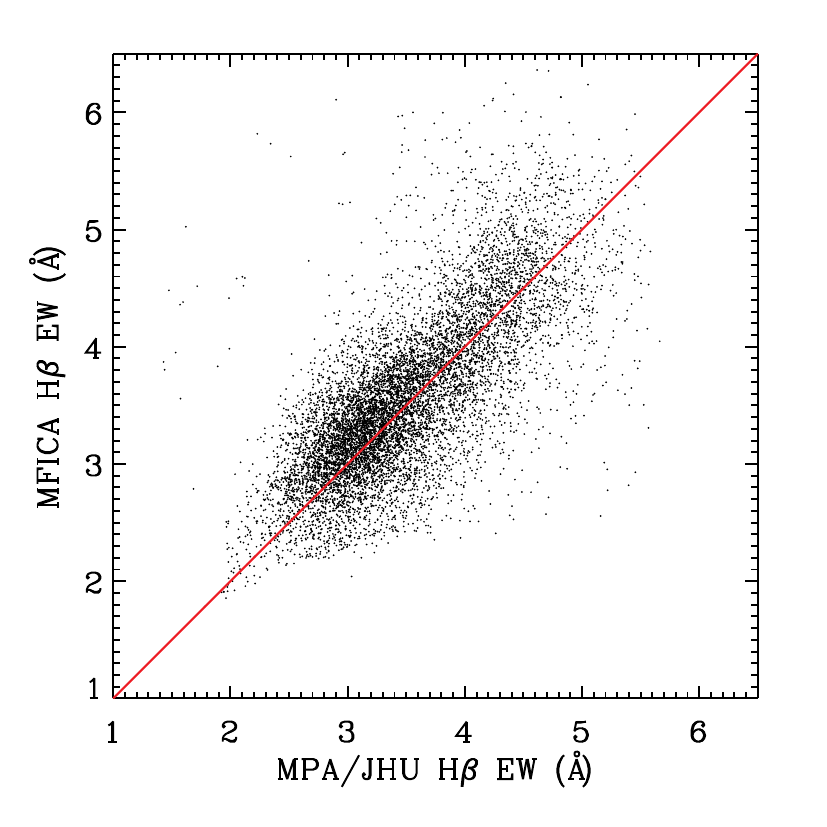}
\caption{Comparison of H$\beta$ EW, defined using the Lick indices, measured from MFICA
  continuum reconstructions and from the MPA/JHU catalogue, for 9932
  emission-line galaxies.}
\label{f:lick}
\end{figure}

Accurate reconstruction of the galaxy continua at locations such as
the Balmer series wavelengths is of particular importance, and is made
more challenging, due to the superposition of strong emission and
absorption features. We test the accuracy of the MFICA Balmer
absorption reconstructions by measuring the H$\beta$ absorption EW,
defined using the Lick indices \citep{Worthey94}, from the continuum
reconstructions for the full sample of $\sim$$10^4$ emission-line
galaxies (see Section~\ref{s:fitting} for a description of the process
of fitting the components to observed spectra). By measuring from the
continuum reconstructions, which by definition do not include the
emission lines, we examine the underlying absorption strength as
inferred by the MFICA analysis. These measurements were compared to
the `model' measurements in the MPA/JHU SDSS
catalogue,\footnote{http://www.mpa-garching.mpg.de/SDSS/} based on
\citet{BC03} stellar population models. The results are
shown in Fig.~\ref{f:lick}, for the 9932 objects that appear in the
MPA/JHU catalogue.

The H$\beta$ measurements are in overall agreement, with a mean offset
of 0.08\,\AA\ and RMS deviation of 0.50\,\AA. For comparison, the mean
offset between the MPA/JHU model measurements and their direct
measurements after subtracting emission lines is $-0.01$\,\AA, with an
RMS deviation of 0.77\,\AA, indicating that the additional scatter
introduced by errors in the MFICA continuum subtraction is smaller
than the measurement error in an individual spectrum. Expressed as a
fraction of the emission line equivalent width, as given in the
MPA/JHU catalogue, the mean offset of the MFICA-based EW is 1.4 per
cent with RMS deviation of 10.2 per cent. We note that the impact on
the H$\beta$ emission-line fluxes will be less than this fraction, as
the emission lines are typically narrower than the absorption lines in
the sample used here. The good agreement between the MFICA
measurements and those based on stellar population models indicates
that any systematic offset in the emission-line fluxes introduced by
the continuum subtraction will be small, and the additional scatter
introduced is also at a low level.

\subsection{Generating emission-line components}
\label{s:emcomps}

The continuum components shown in Fig.~\ref{f:cont_comp} were fitted
to and subtracted from the 727 emission-line galaxy spectra using the
MFICA algorithm with fixed components, again with the wavelength
ranges listed in Table~\ref{t:cont_mask} masked out during the fit.
Preliminary tests showed that applying MFICA directly to the resulting
continuum-subtracted spectra, which cover a range of emission-line
widths, would produce one or more components that were dedicated to
adjusting the widths of the lines. Such components have a large amount
of flux in the wings of the strong emission lines, and little or none
in the centres. To remove the effect of varying
emission-line widths on the MFICA components, the spectra were
convolved with a Gaussian kernel to produce emission lines with a
fixed width of 138\,\kms, corresponding to the broadest lines in the
input sample. Although in doing so we effectively discard information
about the width of the emission lines, this step is necessary in order
to allow the components to be interpreted solely in terms of their
line fluxes, and to limit the number of components required for
accurate reconstructions.

\begin{figure}
\includegraphics[width=\xsizesingle]{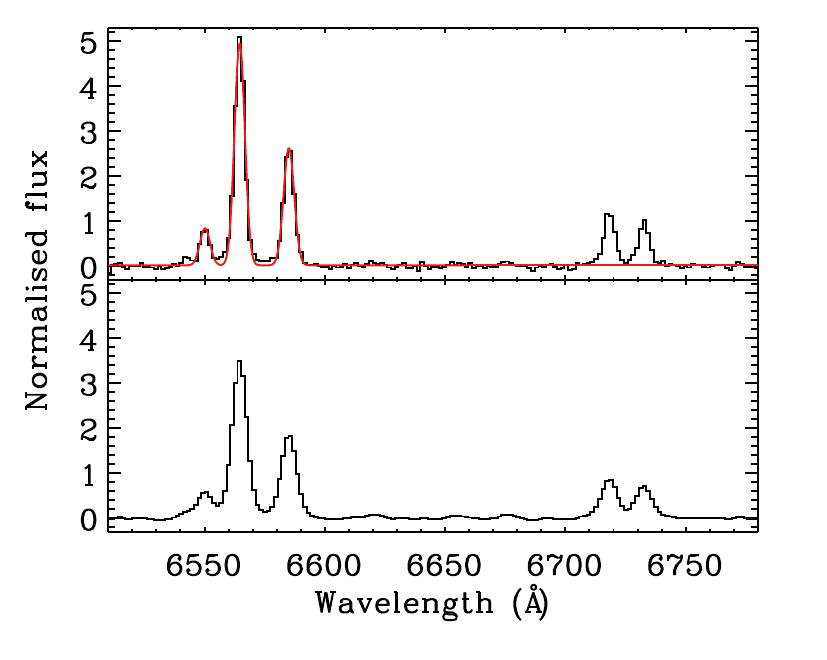}
 \caption{Top: continuum-subtracted narrow emission-line spectrum
   (black) with the Gaussian fit to the H$\alpha$ and \nii\ emission
   lines overplotted (red).  The Gaussian fit has $\sigma_{{\rm
       H}\alpha}=98$\,\kms\ (2.15\,\AA).  Bottom: the same spectrum
   after convolving with a Gaussian kernel of width, $\sigma_{\rm
     K}=97$\,\kms, to give an output width of $\sqrt{(98\,{\rm
       \kms})^2 + (97\,{\rm \kms})^2} = 138$\,\kms.}
 \label{f:blur}
\end{figure}

The width of the Gaussian kernel for each individual galaxy was determined via a simple fit of
three Gaussians to the \nii\,$\lambda\lambda$6550,6585 and H$\alpha$
emission lines.  Each emission-line spectrum was then convolved with a
Gaussian kernel with width, $\sigma_{\rm K} = \sqrt{\left(138\,{\rm
\kms}\right)^2 - \sigma_{{\rm H}\alpha}^2}$, where $\sigma_{{\rm
H}\alpha}$ is the measured H$\alpha$ line width in velocity space, in
order to give all the spectra a 1-$\sigma$ line width of 138\,\kms\ (2
pixels in the SDSS spectra, or 3.02\,\AA\ at the wavelength of
H$\alpha$).  The process is illustrated in Fig.~\ref{f:blur}, which
shows an example continuum-subtracted spectrum before and after the
convolution.  The spectra were then renormalised to have the same total
flux in the emission lines.

The MFICA algorithm itself assumes no knowledge of the specific form
of the components or the mixing matrix, but in the case of
emission-line galaxies we do have some prior knowledge that can be
incorporated into the analysis.  In particular, for some galaxies we
are able to make confident classifications based on the flux ratios of
strong emission lines in a BPT diagram.  \citet{Kauffmann03} defined
star-forming galaxies to be those for which
\begin{equation}
 \log_{10} \left(\frac{{\rm \oiii}\,\lambda5008}{{\rm H}\beta}\right)
 \leqslant \frac{0.61}{\log_{10} \left(\frac{{\rm
       \nii}\,\lambda6585}{{\rm H}\alpha}\right) - 0.05} + 1.3,
\end{equation}
and \citet{Stasinska06} later established that galaxies selected in
this way contain no more than a 3 per cent contribution from an
AGN. Thus, rather than generate components from all emission-line
spectra simultaneously, it is possible to isolate a galaxy subsample
whose spectra are dominated by star formation and use these to derive
a set of `star-formation' components. As described in
Section~\ref{s:loci}, doing so is advantageous when considering the
physical interpretation of the components.

A set of three such components was generated using only the 393
emission-line galaxies out of the sample of 727 that satisfied the \citet{Kauffmann03}
star-forming criterion.  An exponential prior was used, with $\eta=5$.
The increased value of $\eta$ results in a stronger distinction
between high and low flux values, suppressing the flux of the
components in the continuum regions, as expected for emission-line
components.  Increasing $\eta$ to even greater values had negligible
effect on the components. The number of components was chosen by
inspecting the reconstructions of star-forming galaxies produced by
different numbers of components. We found that three components are sufficient to produce
reconstructions of emission-line spectra across the full range of
star-forming galaxies with very high accuracy.

Additional components are necessary to describe the more extended
range of emission-line properties present in AGN spectra. The
additional components were generated using the complete sample of 727
galaxies with the three star-forming components pre-specified and held
fixed. The same prior was used as for the first three emission-line
components. Only a further two components, making a total of five, were
necessary to provide highly accurate reconstructions of essentially
all spectra in the sample. 

\begin{figure*}
\includegraphics[width=\xsizedouble]{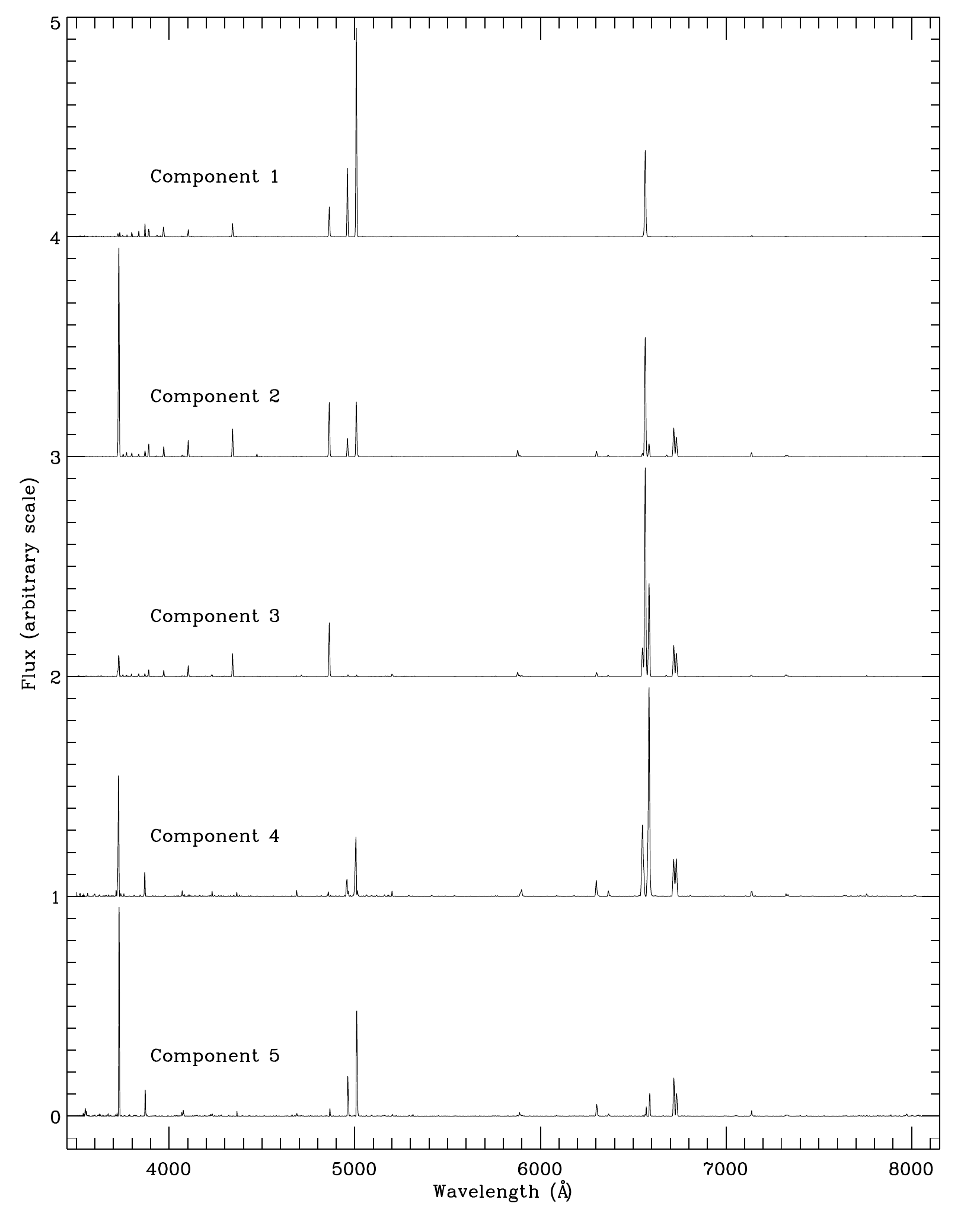}
 \caption{Final set of emission-line components.  Components are
   offset for clarity.  Components 1--3 were generated using
   only the 393 star-forming galaxies; components 4 and 5 used all 727
   galaxies.}
 \label{f:line_comp}
\end{figure*}

\begin{figure*}
\includegraphics[width=\xsizedouble]{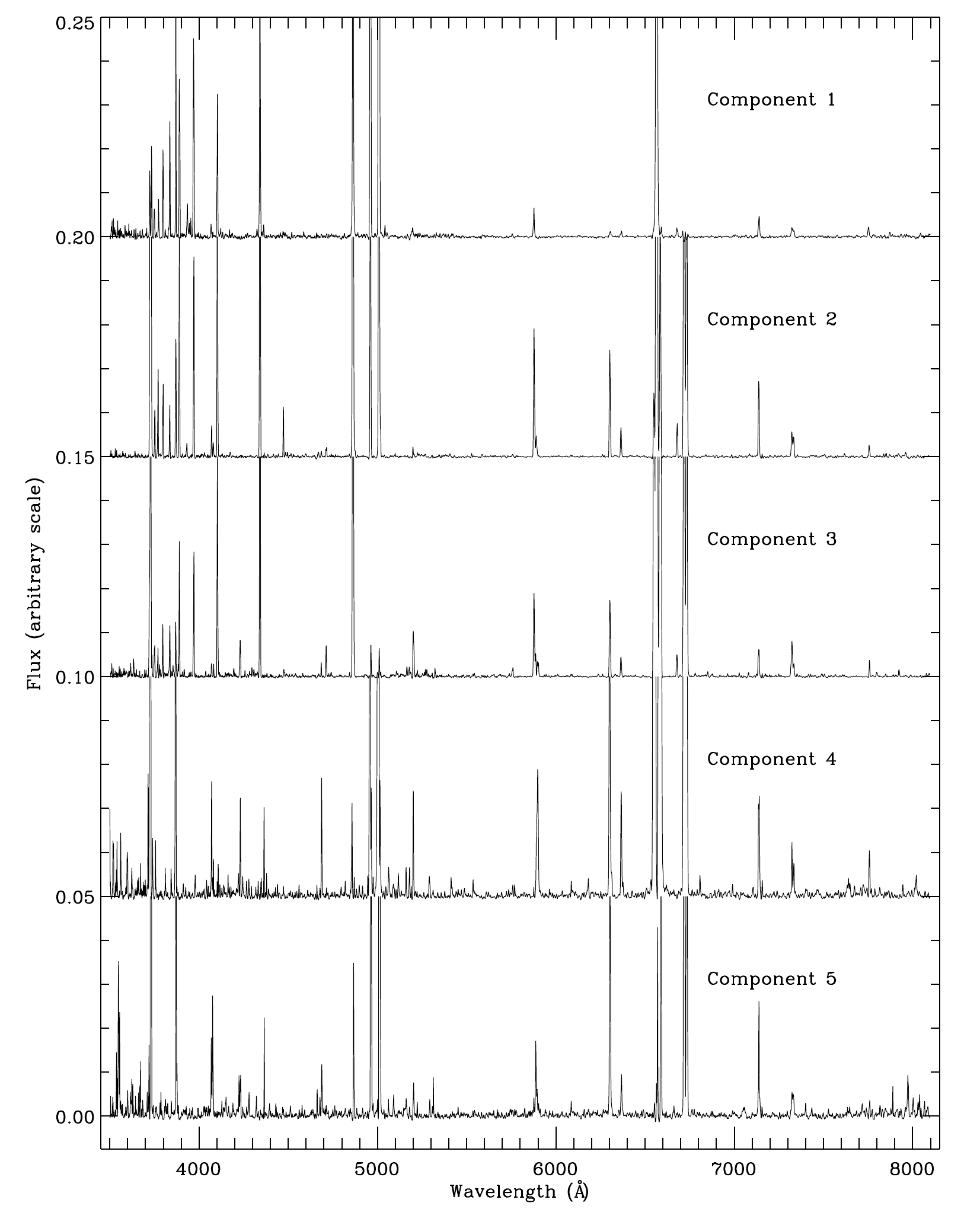}
 \caption{As Fig.~\ref{f:line_comp}, but showing the faint emission
   lines.  Components are offset for clarity.  The ordering of the
   components and the $y$-axis scale are the same as in
   Fig.~\ref{f:line_comp}.}
 \label{f:line_comp_zoom}
\end{figure*}

A detailed examination of the properties of the five individual
MFICA-derived emission-line components was very encouraging with many
weak emission lines evident at high S/N. The only apparent artefact
present in each component was the presence of a very low-level
positive offset, or `signal', of almost constant amplitude independent
of wavelength. The amplitude is extremely small, only
$\simeq$10$^{-3}$ of the emission-line peaks, and well below the
1-$\sigma$ noise in the highest S/N spectra. A simple median-based
filtering scheme, with a window of $\simeq$150\,\AA, was used to
isolate the low-level positive signal, which was then subtracted from
the component, leaving just the emission-line signatures.  The
emission-line components were then normalised such that the sum of
their pixel values is equal to unity (the different normalisation for
the continuum components in Section~\ref{s:concomps} was necessitated
by the presence of components with mean values $\simeq0$).  The resulting set of five
components is shown in Figs.\ \ref{f:line_comp} and
\ref{f:line_comp_zoom}.

The two-stage process, in which the star-formation dominated galaxies
were reproduced first before adding in spectra containing AGN
signatures, greatly aids in the interpretation of the components.  For
instance, we immediately know that any galaxy with significant
contributions from the fourth or fifth components contains spectral
signatures distinct from those of the star-forming galaxies,
indicating the presence of an AGN.  This point is discussed further in
Section~\ref{s:loci}.

An immediately obvious advantage of the MFICA component generation is
that the emission-line components have a much higher S/N than any
individual SDSS spectrum, allowing the identification and measurement
of emission lines that are far too faint to be studied in the
individual input
spectra.  Properties of these faint emission lines are used to
investigate the physical conditions of the emitting gas in
Section~\ref{s:photoionization}.

The five continuum and five emission-line components can be used in
combination to characterise any SDSS narrow emission-line galaxy
spectrum with suitable rest-frame wavelength coverage.  The number of
components required to do so is the same as was found by
\citet{Yip04a} in a related PCA analysis, but the MFICA components
have a number of significant advantages over their PCA counterparts.
In particular, the multi-stage process by which the MFICA components
were constructed provides a clear separation between emission from
different physical sources, while the PCA components each contain a
mixture of continuum and emission-line signals (fig.~20 of Yip et
al. 2004), and have no separation between star formation and AGN
contributions.  The MFICA components are also able to describe a very
broad range of galaxies, while the PCA components struggle to
reconstruct the spectra of extreme emission-line galaxies.  The
separation of the components will prove particularly useful in future
studies exploring the relationship between continuum and emission-line
properties, as it allows the two to be characterised independently of
each other.

\subsection{Fitting components to galaxy spectra}

\label{s:fitting}

Having derived a compact, 10-component, MFICA-generated decomposition
of the carefully selected sub-sample of $\sim$1000 galaxy spectra
the next stage in the analysis is to consider the reconstruction of a
much larger number of SDSS galaxy spectra. The extended sample of
spectra consisted of 10118 emission-line
galaxies (Section~\ref{s:samples}), now with the full range of
spectrum S/N ($15.0 \leqslant {\tt SN\_R} < 30.0$) and a slightly
broader range of H$\alpha$ emission-line width
($2.0\,{\text{\AA}}\leqslant \sigma_{{\rm H}\alpha} <
3.5\,{\text{\AA}}$).

\begin{figure*}
\includegraphics[width=\xsizedouble]{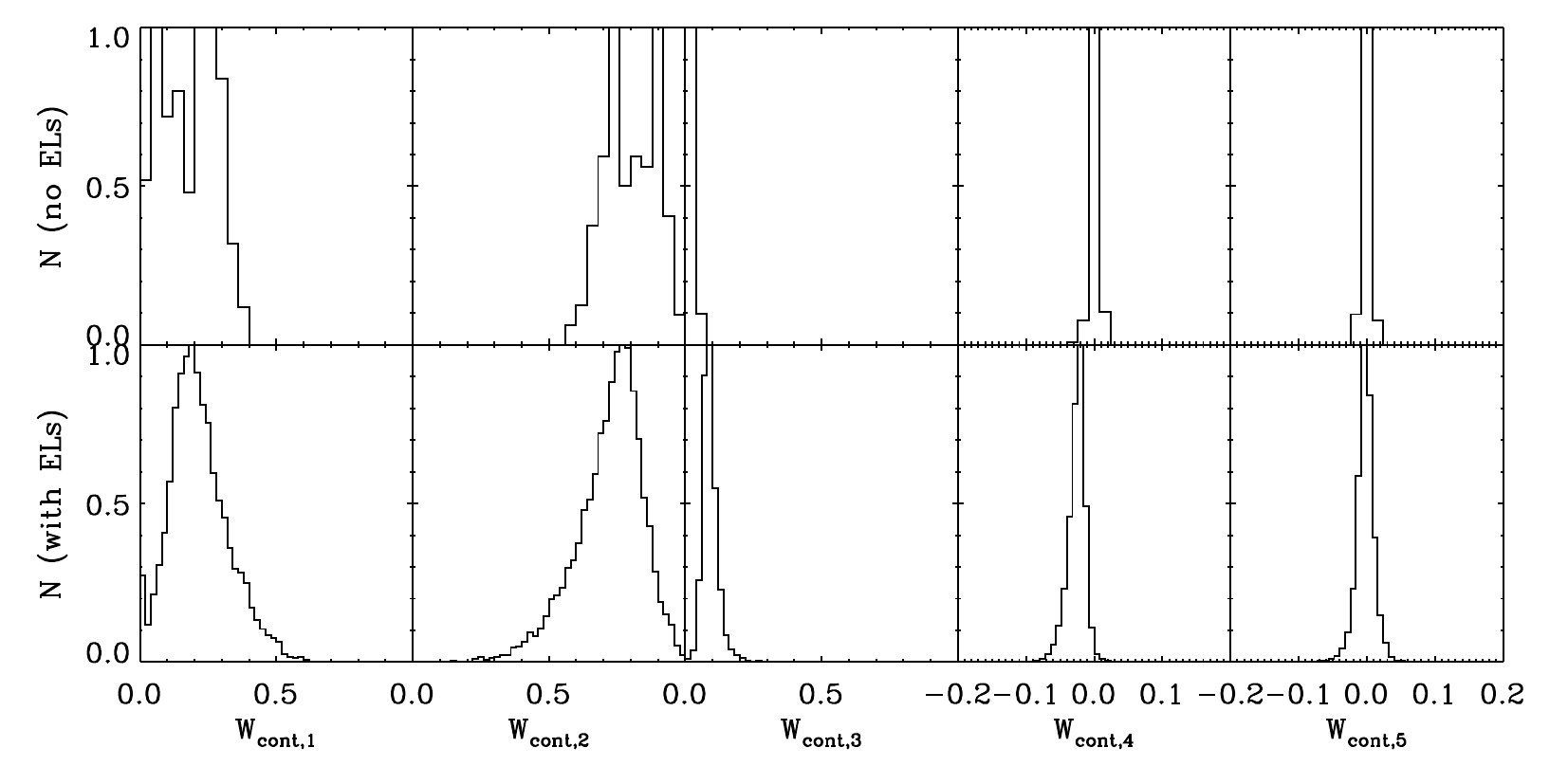}
 \caption{Distributions of the individual continuum component
   weights, each normalised to their maximum values. The upper panels
   are for the 170 galaxies without emission lines (ELs), which formed the
   bulk of the sample from which the first three continuum components
   were derived. The lower panels are for the full sample of 10118
   emission-line galaxies.}
 \label{f:w_cont_hist}
\end{figure*}

\begin{figure*}
\includegraphics[width=\xsizedouble]{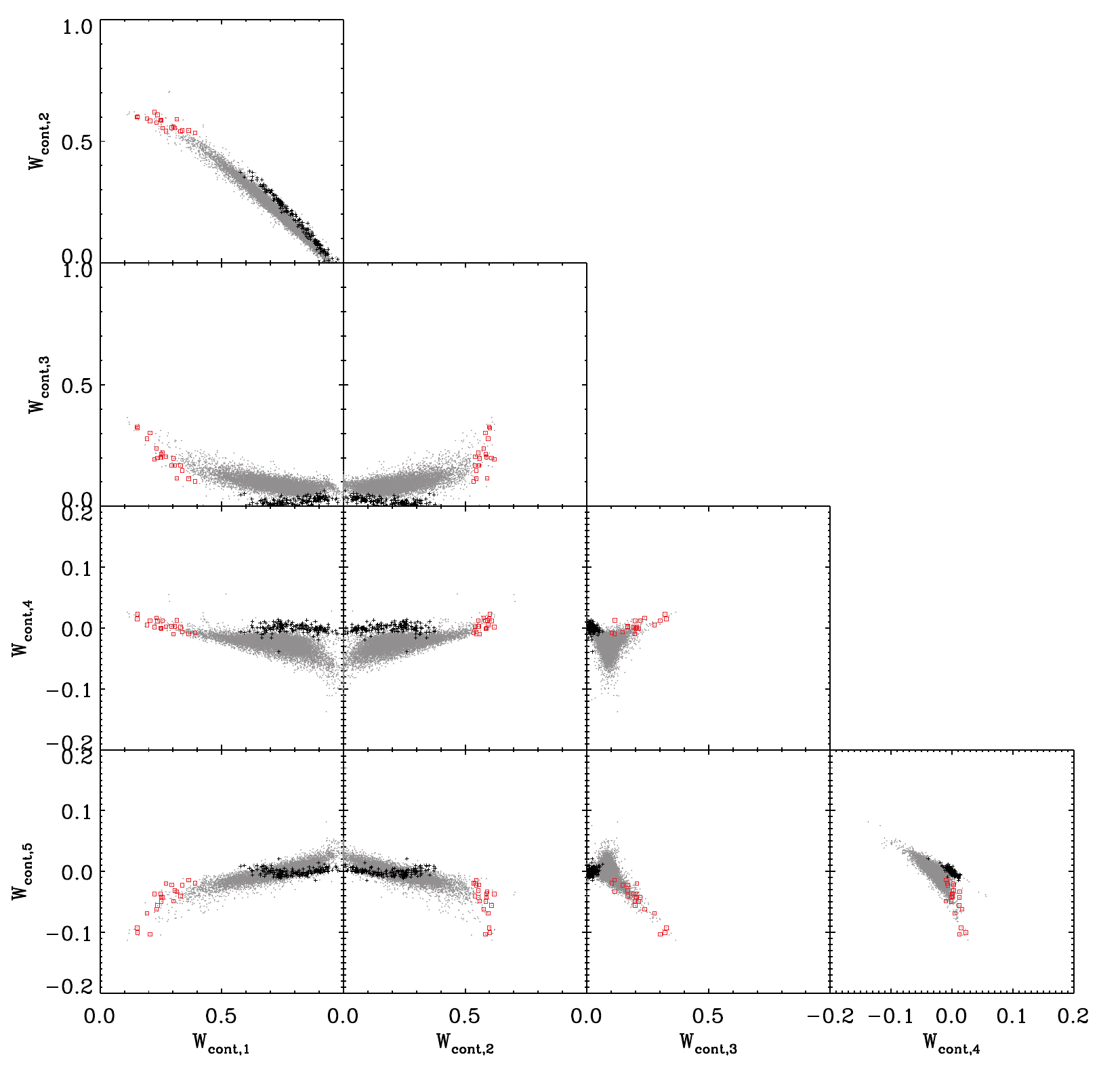}
 \caption{Distributions of continuum component weights.
   The larger black points are for the 170 galaxies without emission
   lines; red squares are for the 20 emission-line galaxies used when
   generating continuum components 1--3; smaller grey points are for
   the full sample of 10118 emission-line galaxies.}
 \label{f:w_cont}
\end{figure*}

To fit the continuum components to each galaxy a $\chi^2$ minimization
was performed, using the mask defined in Table~\ref{t:cont_mask}, with
the weights for the first three components constrained to be positive.
In the $\chi^2$ minimization, and the steps described below, the SDSS noise array was
used rather than the (scalar) noise covariance, ${\bm \Sigma}$, defined
as part of MFICA itself in Section~\ref{s:mfica}. An accurate redshift
for the emission lines is particularly important
due to their rapidly varying nature as a function of wavelength. After
subtracting the continuum, the redshift of the emission lines
was remeasured from the H$\alpha$ line, by fitting single Gaussians to
\nii\,$\lambda\lambda$6550,6585 and H$\alpha$.  The observed spectra
were then adjusted to the new redshift, and the continuum components
were refitted and subtracted.  The resulting weights were normalised
such that they sum to unity; in the following we denote the normalised
weights by $W_{{\rm cont},i}$. The distributions of the continuum
weights are shown in Figs.\
\ref{f:w_cont_hist} and \ref{f:w_cont}. An alternative scheme was also tested
in which the redshift of the continuum components was left as a free
parameter in the fit, but doing so did not give any further
improvement in the quality of the continuum subtraction.  It is
important to note that the MFICA-based continuum subtraction
incorporates reconstructions of the stellar absorption features,
allowing accurate subtractions even where these features are
coincident with emission lines, such as at H$\beta$.

Fig.~\ref{f:w_cont} shows some intriguing differences between galaxies
with and without emission lines. Very low values of
$W_{\mathrm{cont},3}$ in the galaxies without emission lines suggest a
lack of recent star formation, as expected. These galaxies also have
very small values of $W_{\mathrm{cont},4}$ and $W_{\mathrm{cont},5}$;
in Section~\ref{s:vespa} we show that this corresponds to the lowest
observed levels of dust reddening. However, as these galaxies were
themselves used in generating the continuum components, care must be
taken when comparing their weights to those of the full sample. As
expected from the way in which they were selected, the 20
emission-line galaxies used in generating continuum components 1--3
show extremely high values of 
$W_{\mathrm{cont},2}$, and often high values of $W_{\mathrm{cont},3}$
as well, indicating high levels of recent or ongoing star formation. The
continuum weights of the emission-line galaxies are examined in more
detail in Section~\ref{s:vespa}.

The range of emission-line width in galaxies used to define the sample
from which the MFICA components were derived was deliberately
constrained to have an upper limit of $\sigma_{{\rm H}\alpha} <
3.02$\,\AA. The rationale was to retain the maximum information,
evident at relatively high spectral resolution, in the MFICA
emission-line components. The SDSS-selected galaxy population as a
whole includes objects with significantly broader emission-line
velocity widths, although the percentage of such objects is small; for
galaxies that do not possess a visible broad emission-line AGN
component approximately 3 per cent possess $\sigma_{{\rm H}\alpha} >
3.5$\,\AA \ and well under 0.5 per cent have $\sigma_{{\rm H}\alpha} >
5.0$\,\AA.

\begin{figure*}
\includegraphics[width=\xsizedouble]{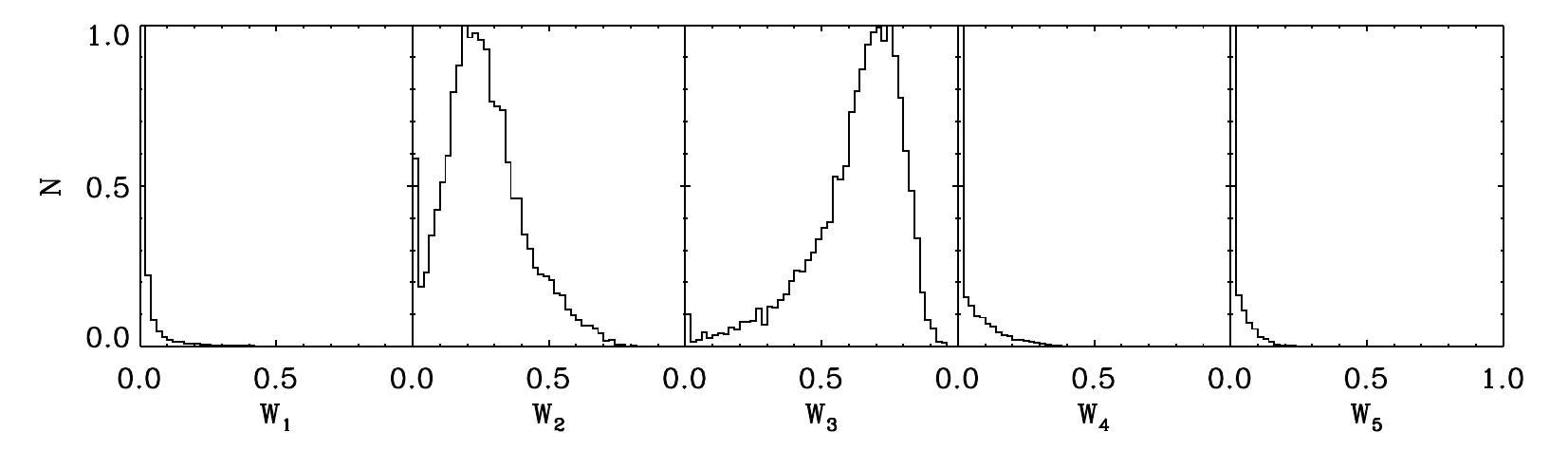}
 \caption{Distributions of the individual emission-line component
   weights, each normalised to their maximum values.}
 \label{f:w_line_hist}
\end{figure*}

\begin{figure*}
\includegraphics[width=\xsizedouble]{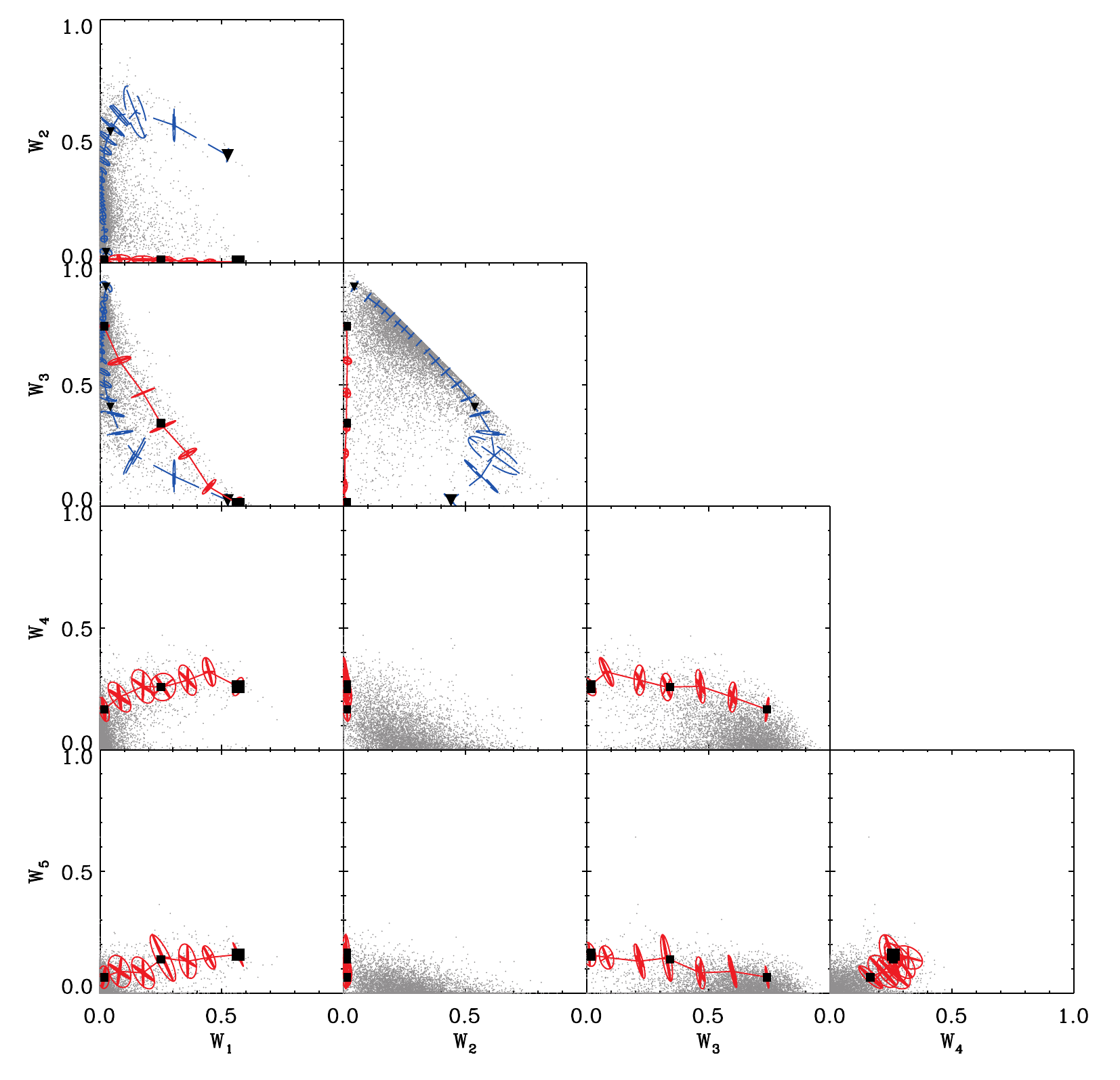}
 \caption{Distributions of emission-line component weights.
   Overplotted are the loci for SF (dashed blue) and AGN (solid red).
   The positions of the representative spectra are also shown for SF
   (triangles) and AGN (squares); larger symbols represent s2 and a2,
   i.e.\ the spectra with the highest \oiii$/$H$\beta$.}
 \label{f:w_line}
\end{figure*}

It is essential that the emission-line widths of individual galaxies
and the MFICA components match very closely in order for an accurate
decomposition to be obtained. For galaxies with H$\alpha$ width
$\sigma_{{\rm H}\alpha} < 138$\,\kms (3.02\,\AA), the spectra were
convolved with a Gaussian kernel to produce a spectrum with
$\sigma_{{\rm H}\alpha} = 138$\,\kms, following the same procedure as
described in Section~\ref{s:emcomps}. Where a galaxy possessed an
H$\alpha$ width exceeding 138\,\kms, the MFICA emission-line
components were convolved with a Gaussian kernel to produce components
with the same emission-line width as in the galaxy. With the galaxy
and component emission-line widths made equal the emission-line
components were then fitted to the spectra, using a $\chi^2$
minimization with all component weights constrained to be positive.
The weights were normalised such that they sum to unity.  The
distributions of emission-line component weights, $W_i$, are shown in
Figs.\ \ref{f:w_line_hist} and \ref{f:w_line}.  The median values of
the formal 1-$\sigma$
uncertainties are 0.004, 0.008, 0.009, 0.006 and 0.009 in each of the
weights, respectively.

\begin{figure*}
\includegraphics[width=\xsizedouble]{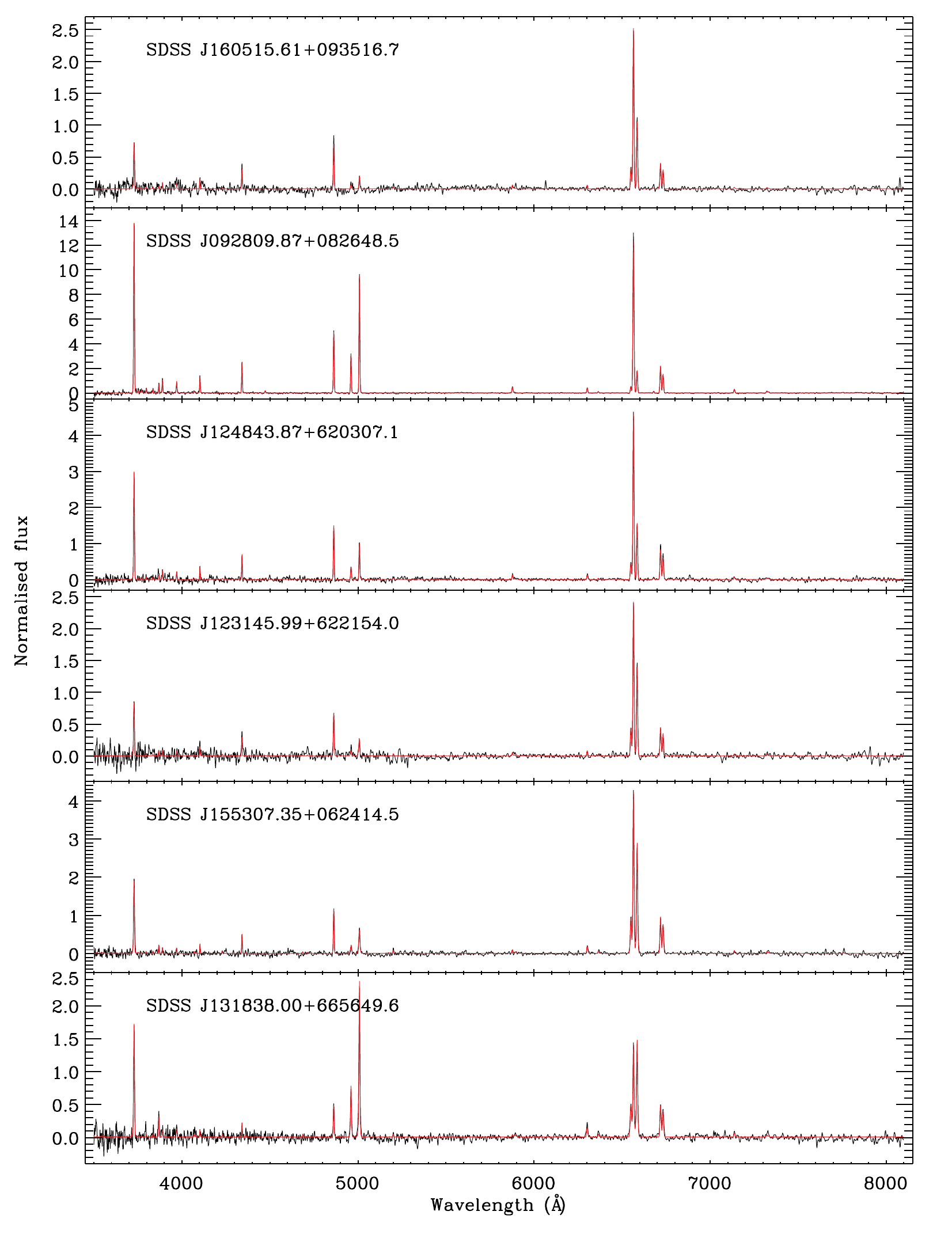}
 \caption{Example MFICA reconstructions of pure star-forming galaxies
   (top three panels) and those with AGN contributions (bottom three
   panels).  In each panel the continuum-subtracted spectrum is
   plotted in black, and the MFICA reconstruction in red. The object
   name is given in each panel.}
 \label{f:example_fit}
\end{figure*}

The emission-line components allow highly accurate descriptions of
galaxies across the entire populated area in the \oiii$/$H$\beta$
vs.\ \nii$/$H$\alpha$ plane.  Median absolute deviation fractional
errors in the reconstructed fluxes are 1.8 per cent in H$\alpha$, 7.6
per cent in H$\beta$ and 12.7 per cent in \oiii\,$\lambda$5008,
corresponding, given the S/N of the spectra, to 1.3$\sigma$,
1.8$\sigma$ and 1.2$\sigma$, respectively. Assuming Gaussian
  errors in the measurements of observed flux, perfect reconstructions
  would result in a median absolute deviation of 0.67$\sigma$ due to
  observational noise.  Subtracting the reconstructions from the
emission-line spectra removes 93 per cent of the excess RMS in the
emission lines, relative to that in the continuum. Example reconstructions are shown in
Fig.~\ref{f:example_fit} for both pure star-forming galaxies and those
with AGN, illustrating the accuracy of the fits for a range of galaxy
properties.

As the MFICA components, and hence the reconstructions, have
considerably higher S/N than the individual SDSS spectra, flux
measurements can be made from a reconstruction for emission lines that
are too weak to be measured in the corresponding observed
spectrum. However, the form of the reconstruction is itself dominated
by the strong emission lines, so any measurement of weak line fluxes
in an individual galaxy should be viewed as a prediction based on the
properties of the stronger lines. This prediction is, in essence, an
average over a number of galaxies with similar properties in their
strong emission lines.

\subsection{Definition of star formation and active galactic nuclei
  loci}

\label{s:loci}

A primary goal of the application of the MFICA-component decomposition
of the galaxy spectra is to generate a quantitative estimate of the
contribution of both star-formation-related processes and any AGN
that may be present to the observed spectrum.  The two-stage process
used to derive the emission-line components already gives important
information about the interpretation of the component weights, as any
galaxy with significant contributions from the fourth or fifth
components has a spectrum that cannot be due to star formation alone,
strongly suggesting the presence of an AGN. However, this information
alone is insufficient to quantify the star-formation and AGN
contributions to all galaxies, or to extract the spectral signature of
these individual contributions, and a more precise MFICA-based
definition of SF and AGN is required.

Star formation and AGN each produce observed emission-line spectra
that display a range of properties and even `pure' examples of each do
not correspond to individual points in the five-dimensional space of
MFICA emission-line component weights.  Instead, more extended regions
of the space are occupied by objects whose spectra arise solely from
star-formation-related processes or manifestations of an AGN. In the
case of star formation for example, the emission-line spectrum evolves
significantly as a function of the age since a starburst and factors
such as the initial mass function (IMF) of the burst and the
metallicity of the gas may also contribute to the diversity of
spectral properties.

The form of the distribution of star-formation- and AGN-dominated
spectra in the classical BPT diagrams, together with visual inspection
of the location of the galaxy spectra in the five-dimensional space of
the MFICA component weights, strongly suggest that the `pure' examples
are restricted to limited regions of the space of MFICA weights. A
natural choice was therefore to parametrize these regions as
multi-dimensional loci, using the algorithm presented by \citet{NY97}.
The algorithm uses an iterative procedure to define a set of locus
points that follow the centre of an extended distribution of data
points.  At each locus point the distribution of data points around
the locus is described by an ellipse (or higher-dimensional
ellipsoid).

The algorithm was implemented with $N_{\sigma_{\rm spacing}}=3.0$, no
maximum distance for inclusion in the locus (see \citet{NY97} for
definitions of these parameters) and an additional requirement that at
least 20 data points must exist between two locus points for an extra
locus point to be inserted.

The star-formation locus was defined using the 5519 objects that
satisfied the \citet{Kauffmann03} star-forming definition.  As these objects are each
expected to contain no more than a 3 per cent contribution to their
emission-line flux from an AGN, they can be used to explore the range
of component weights corresponding to star-formation-dominated
spectra.  The locus was assumed to carry zero weight from the final
two emission-line components and is defined only in the
three-dimensional space of the first three component weights. The
resulting locus is plotted in blue in Fig.~\ref{f:w_line}.  Moving
along the length of the locus is approximately equivalent to moving
along the star-formation `wing' of the BPT diagram.

The corresponding AGN locus is more challenging to define, as objects
selected from BPT-based definitions may contain substantial
star-formation contributions.  However, it can be seen in
Fig.~\ref{f:w_line} that the SF locus has non-zero contributions from
the second component throughout its length, other than at the extreme
end, where the weight of the third component tends to unity.  As such,
those objects with minimal contributions from the second component can
be assumed to be pure, or nearly-pure, AGN\@.  The AGN locus was
defined in all five dimensions from the 379 objects with weights $W_2
\leqslant 0.05$ and $W_4 + W_5 \geqslant 0.18$.  The second criterion
ensures a clean separation of the star-formation and AGN loci;
relaxing the criterion gives no improvement in the ability of the loci
to describe observed galaxies, but greatly increases the degeneracies
in the decomposition of objects into star-formation and AGN
contributions, suggesting that the objects excluded by this criterion
are primarily composite objects.

The AGN locus is plotted in red in Fig.~\ref{f:w_line}.  For each of
the star-formation and AGN loci the 1-$\sigma$ radii generated by the
\citet{NY97} algorithm were multiplied by 1.5 to more fully encompass
the contributing objects.  The resulting ellipses, defining the
transverse width of the loci at each point along their length, are
included in Fig.~\ref{f:w_line}.

\section{Results and discussion}

The MFICA components derived in Section~\ref{s:method} provide new
avenues for the investigation of galaxy properties. The high S/N
reconstructions that result from the components can be examined to
discern the detailed physical properties of the corresponding emission
regions. Additionally, the set of weights measured for any individual
galaxy acts as a compact representation of its observational
properties, and hence the distributions of these weights carry
information about the distributions of various properties within the
galaxy sample under investigation. Although we defer a full
exploration of these possibilities to future papers, we here present a
brief illustration of the correlations between the MFICA continuum
weights and previously-measured star formation histories, as well as
preliminary results from photoionization modelling of a range of
emission-line spectra.

\label{s:results}

\subsection{Comparison with VESPA results}

\label{s:vespa}

The clear identification of the three positive continuum components
with old, intermediate and young stellar populations, respectively,
allows the weights of these components within any individual galaxy to
be used as a crude measure of the galaxy's star formation history
(SFH).  A full development of such a technique is beyond the scope
of this paper, but we present here a comparison of our results with
those from a recent catalogue of SFHs, in order to demonstrate the
success of the MFICA algorithm in identifying a variety of different
stellar populations.

The VESPA algorithm \citep{Tojeiro07} derives SFHs by fitting
combinations of simple stellar population models to observed galaxy
spectra.  The metallicity of each stellar population is also derived,
along with a measurement of the overall dust content.  The number of
free parameters is varied according to the S/N and other properties of
each individual spectrum, ensuring a robust determination of the SFH
at the level of precision that is warranted by the data.
\citet{Tojeiro09} present a catalogue of VESPA-derived SFHs for the
SDSS, from which we draw our comparison data.  Of the 10118 galaxies
in the emission-line galaxy sample, VESPA data were available for
9113; we retrieved the data for these galaxies derived using the
\citet{BC03} stellar population models and a single-parameter dust
model.  The extinction law is based on the mixed slab model of
\citet{CF00}, characterised by the optical depth at 5500\,\AA,
$\tau_V^{\rm ISM}$.\footnote{\citet{Tojeiro09} also present results
  using a two-parameter dust model that allows a higher level of dust
  in birth clouds.  For simplicity we restrict our comparison to the
  single-parameter dust model here.}

\begin{figure}
\includegraphics[width=\xsizesingle]{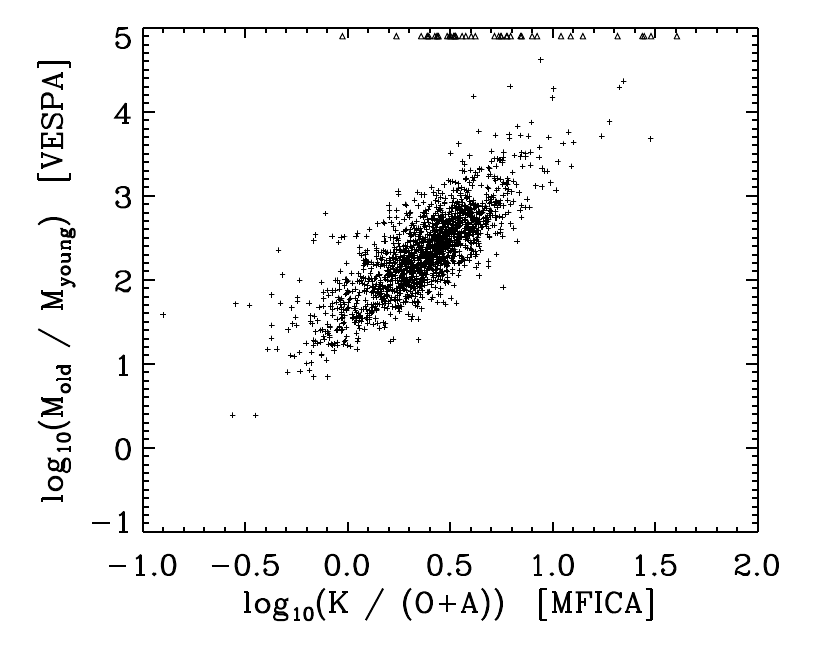}
 \caption{VESPA-derived ratio of old ($>$0.42\,Gyr) to young
   ($<$0.42\,Gyr) stellar masses, vs.\ ratio of MFICA weights for old
   (K-giant dominated) and young (O- and A-star dominated) stellar
   populations.  Objects beyond the limits of the figure -- typically
   because $M_{\rm young}=0$ -- are marked with triangles.
   Only those galaxies with $\tau_V^{\rm ISM}\leq0.75$ are shown.}
 \label{f:vespa_mass_cut}
\end{figure}

A direct comparison of the MFICA and VESPA results is shown in
Fig.~\ref{f:vespa_mass_cut}.  In this figure the $x$-axis shows the
MFICA component weight for the old stellar population (K-giant
dominated) divided by the summed weights of the younger populations
(O- and A-star dominated).  The $y$-axis shows the VESPA-derived ratio
of old ($>$0.42\,Gyr) to young ($<$0.42\,Gyr) stellar masses.  The
division at 0.42\,Gyr splits the logarithmically-spaced VESPA age bins
in half, and is approximately equal to the main sequence lifetime of
an A-star.  Only galaxies with $\tau_V^{\rm ISM}\leq0.75$ are shown.
Even though the figure compares a ratio of masses to a ratio of
luminosities, there is a very strong correlation between the two.

\begin{figure}
\includegraphics[width=\xsizesingle]{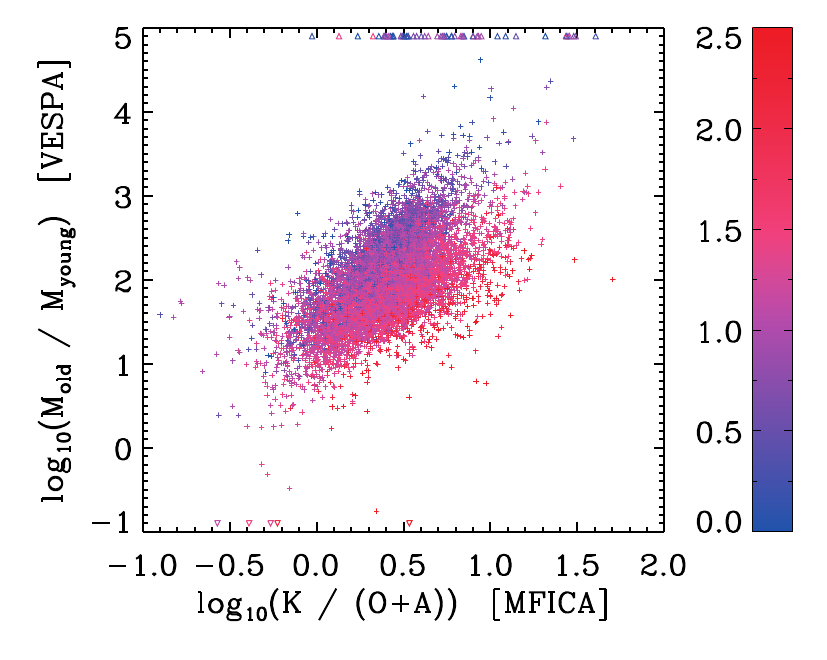}
 \caption{As Fig.~\ref{f:vespa_mass_cut}, but showing galaxies
   with all values of $\tau_V^{\rm ISM}$, as denoted by the colour
   scale.  There is a systematic increase in $\tau_V^{\rm ISM}$
   towards the bottom-right corner.}
 \label{f:vespa_mass}
\end{figure}

The dust reddening explicitly included in the VESPA fit allows it to
distinguish between a young dust-reddened stellar population, and an
older population with little or no dust, which may have a similar
overall spectral shape.  In contrast, the MFICA fit does not
explicitly include the effect of dust, so dust reddening is accounted
for by increasing the contribution from an old, red, stellar
population.  This effect can be seen in Fig.~\ref{f:vespa_mass}, which
shows the same data as Fig.~\ref{f:vespa_mass_cut} but for all values
of $\tau_V^{\rm ISM}$, denoted by the colour scale.  The different
methods by which the two algorithms account for dust results in the
systematic increase in $\tau_V^{\rm ISM}$ that can be seen when
moving to higher $\log_{10}({\rm K}/({\rm O+A}))$ at any fixed
$\log_{10}(M_{\rm old}/M_{\rm young})$.

\begin{figure}
\includegraphics[width=\xsizesingle]{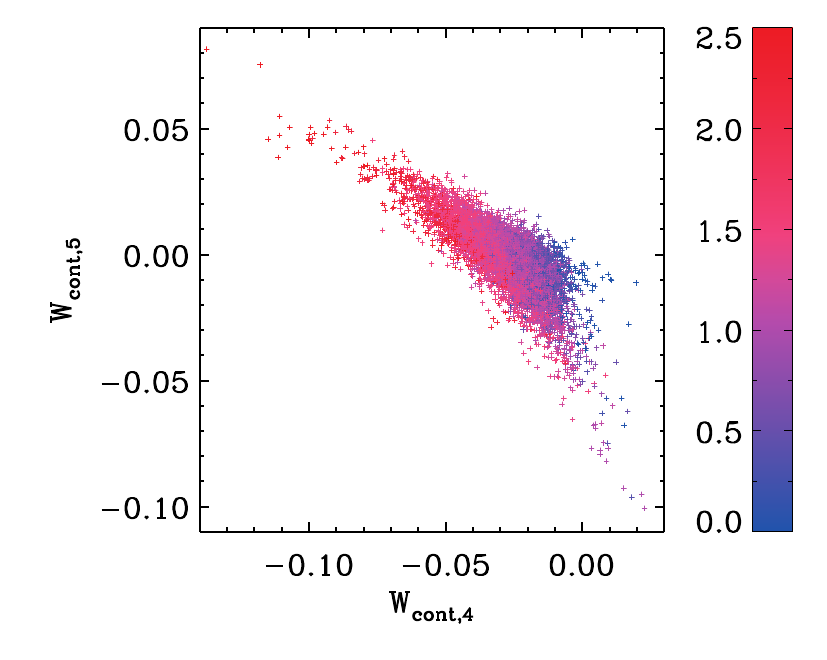}
 \caption{MFICA weights for the two continuum `adjustment'
   components.  Points are colour-coded according to their
   VESPA-derived $\tau_V^{\rm ISM}$ value, as in
   Fig.~\ref{f:vespa_mass}; there is a strong correspondence between
   $\tau_V^{\rm ISM}$ and position in the $W_{{\rm cont},4}$--$W_{{\rm
     cont},5}$ plane.}
 \label{f:vespa_mass_w45}
\end{figure}

Although the MFICA algorithm does not explicitly account for dust, it
is still possible to distinguish between different levels of dust
reddening by considering the full 5-dimensional distribution of
component weights.  As an illustration of this potential,
Fig.~\ref{f:vespa_mass_w45} shows the MFICA weights for the two
`adjustment' components, with the value of $\tau_V^{\rm ISM}$ again
denoted by the colour scale.  It is clear that the position of an
individual galaxy in the $W_{{\rm cont},4}$--$W_{{\rm cont},5}$ plane
can be used to predict its level of dust reddening, and hence, when
combined with the other component weights, the nature of its stellar
populations.

Although we do not present a calibrated method for deriving SFHs, the
strong correlations between the VESPA results and the MFICA continuum
component weights illustrate the success of the MFICA technique in
identifying and quantifying the contributions from stellar populations
of different ages.  Future studies will unlock the full potential of
this technique by establishing the exact nature of the correspondence
between the SFH of a galaxy and its derived MFICA weights.

\subsection{Physical interpretation of MFICA loci}

\label{s:photoionization}

\subsubsection{Emission line strengths}

\begin{figure*}
\includegraphics[width=\xsizedouble]{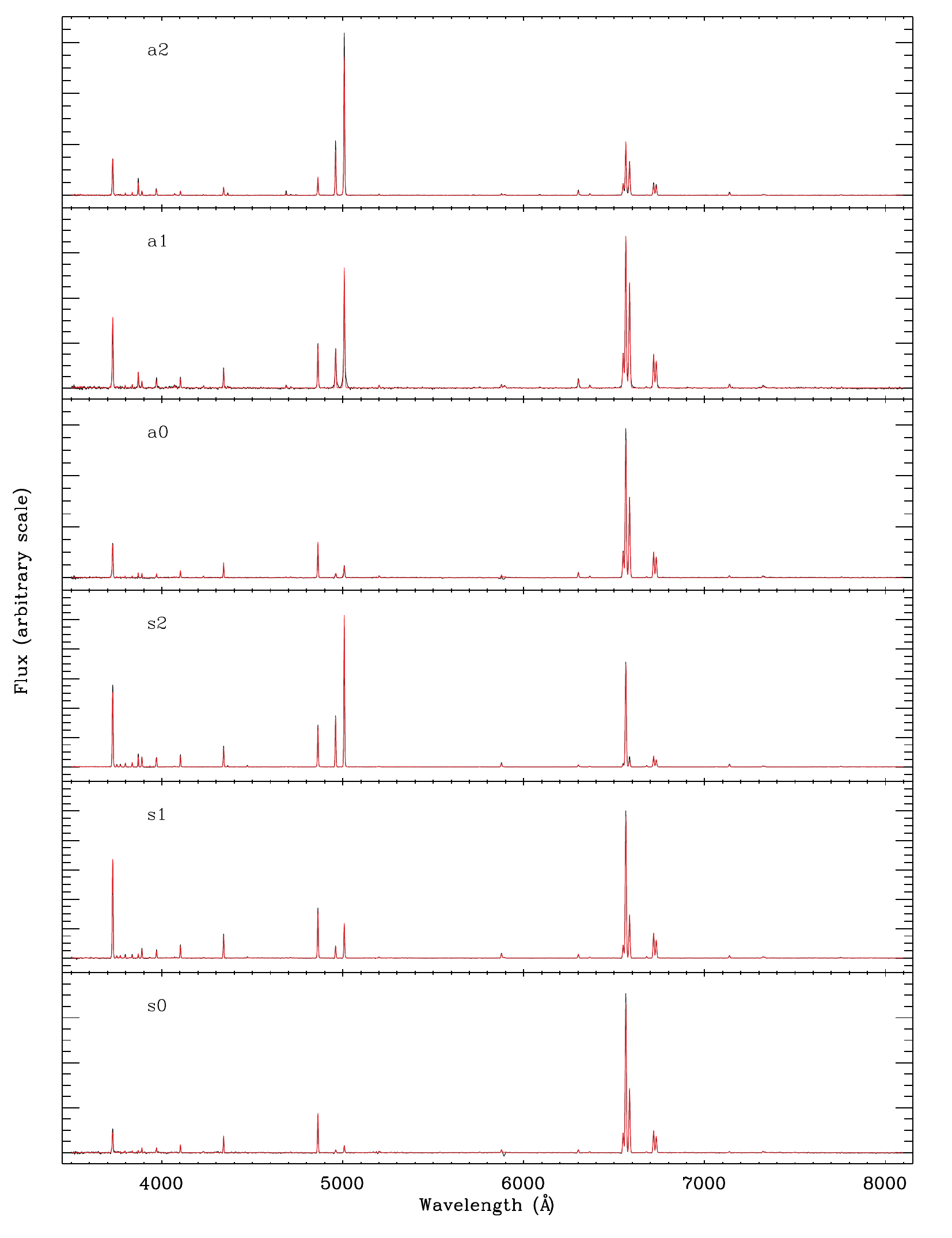}
 \caption{Reconstructed MFICA spectra from a range of positions within
   the AGN (top three panels) and SF (bottom three panels) loci.  In
   each panel the MFICA reconstruction is in plotted in red, and a
   composite of the continuum-subtracted spectra of 50 observed
   galaxies with similar MFICA weights is in black.}
 \label{f:templates}
\end{figure*}

In order to explore the physical properties of galaxies within the AGN
and SF loci, we generated a series of three reconstructed MFICA
spectra lying along the extent of each locus.  In the AGN locus, point
a0 falls at a position very near to the star-forming sequence, point
a2 is the opposite end of the locus and corresponds to the extreme AGN
case, while point a1 is midway along the locus.  Similarly, points s0,
s1, s2 correspond to increasingly high ionization levels along the SF
locus.  The six reconstructed spectra are shown in
Fig.~\ref{f:templates}, along with composite spectra of observed
galaxies with similar MFICA weights.  Their component weights are
included in Fig.~\ref{f:w_line}.  Table~\ref{t:lineratio} lists
emission line strengths measured directly from the reconstructed MFICA
spectra, and also values dereddened (using a standard Galactic
reddening curve) so that $I({\rm H}\alpha)/I({\rm H}\beta) = 2.86$.

\begin{table*}
  \centering
  \caption{Measured and dereddened emission line strengths for
    reconstructed spectra.  Measurements are given relative to H$\beta$.}
  \label{t:lineratio}
  \begin{tabular}{cr|r@{.}lr@{.}lr@{.}lr@{.}lr@{.}lr@{.}l|r@{.}lr@{.}lr@{.}lr@{.}lr@{.}lr@{.}l}
    \hline
\multirow{2}{*}{Ion} & \multirow{2}{*}{$\lambda_{\rm vac}$} & 
\multicolumn{12}{|c|}{Measured} & \multicolumn{12}{|c}{Dereddened} \\
 & & \multicolumn{2}{|c}{a0} & \multicolumn{2}{c}{a1} &
\multicolumn{2}{c}{a2} & \multicolumn{2}{c}{s0} &
\multicolumn{2}{c}{s1} & \multicolumn{2}{c|}{s2} &
\multicolumn{2}{|c}{a0} & \multicolumn{2}{c}{a1} &
\multicolumn{2}{c}{a2} & \multicolumn{2}{c}{s0} &
\multicolumn{2}{c}{s1} & \multicolumn{2}{c}{s2} \\
\hline
\oii & 
3729 & 
0&98 & 
1&79 & 
2&21 & 
0&53 & 
2&13 & 
1&99 & 
1&80 & 
3&02 & 
3&05 & 
0&97 & 
3&00 & 
2&38 \\
\hi & 
3751 & 
0&01 & 
0&01 & 
0&02 & 
0&02 & 
0&02 & 
0&03 & 
0&03 & 
0&02 & 
0&03 & 
0&03 & 
0&03 & 
0&04 \\
\hi & 
3772 & 
0&00 & 
0&01 & 
0&02 & 
0&01 & 
0&02 & 
0&04 & 
0&01 & 
0&01 & 
0&03 & 
0&01 & 
0&03 & 
0&04 \\
\hi & 
3799 & 
0&02 & 
0&04 & 
0&10 & 
0&02 & 
0&03 & 
0&07 & 
0&03 & 
0&07 & 
0&13 & 
0&04 & 
0&05 & 
0&09 \\
\hi & 
3837 & 
0&03 & 
0&05 & 
0&12 & 
0&03 & 
0&03 & 
0&08 & 
0&05 & 
0&09 & 
0&16 & 
0&05 & 
0&04 & 
0&09 \\
\neiii & 
3870 & 
0&11 & 
0&31 & 
0&55 & 
0&04 & 
0&07 & 
0&18 & 
0&20 & 
0&49 & 
0&74 & 
0&07 & 
0&10 & 
0&22 \\
\hi & 
3890 & 
0&08 & 
0&13 & 
0&23 & 
0&09 & 
0&13 & 
0&21 & 
0&14 & 
0&20 & 
0&31 & 
0&15 & 
0&18 & 
0&24 \\
\hi & 
3971 & 
0&08 & 
0&14 & 
0&25 & 
0&09 & 
0&12 & 
0&20 & 
0&14 & 
0&21 & 
0&33 & 
0&14 & 
0&16 & 
0&23 \\
\sii & 
4070 & 
0&02 & 
0&04 & 
0&06 & 
0&01 & 
0&01 & 
0&02 & 
0&03 & 
0&06 & 
0&07 & 
0&02 & 
0&02 & 
0&02 \\
\hi & 
4103 & 
0&17 & 
0&18 & 
0&20 & 
0&17 & 
0&21 & 
0&22 & 
0&26 & 
0&26 & 
0&25 & 
0&27 & 
0&26 & 
0&25 \\
\fev ? & 
4231 & 
0&04 & 
0&04 & 
0&03 & 
0&03 & 
0&02 & 
0&00 & 
0&06 & 
0&06 & 
0&04 & 
0&04 & 
0&02 & 
0&00 \\
\hi & 
4342 & 
0&36 & 
0&37 & 
0&39 & 
0&36 & 
0&41 & 
0&43 & 
0&48 & 
0&47 & 
0&46 & 
0&49 & 
0&48 & 
0&47 \\
\oiii & 
4364 & 
0&01 & 
0&02 & 
0&04 & 
0&00 & 
0&00 & 
0&01 & 
0&01 & 
0&03 & 
0&05 & 
0&00 & 
0&00 & 
0&01 \\
\hei & 
4473 & 
0&01 & 
0&01 & 
0&01 & 
0&00 & 
0&02 & 
0&02 & 
0&01 & 
0&01 & 
0&01 & 
0&01 & 
0&02 & 
0&02 \\
\heii & 
4687 & 
0&02 & 
0&04 & 
0&05 & 
0&01 & 
0&00 & 
0&00 & 
0&02 & 
0&04 & 
0&05 & 
0&01 & 
0&01 & 
0&00 \\
\ariv & 
4713 & 
0&02 & 
0&02 & 
0&01 & 
0&02 & 
0&01 & 
0&01 & 
0&02 & 
0&02 & 
0&01 & 
0&02 & 
0&02 & 
0&01 \\
\hi & 
4863 & 
1&00 & 
1&00 & 
1&00 & 
1&00 & 
1&00 & 
1&00 & 
1&00 & 
1&00 & 
1&00 & 
1&00 & 
1&00 & 
1&00 \\
\oiii & 
4960 & 
0&17 & 
1&04 & 
2&65 & 
0&07 & 
0&25 & 
1&26 & 
0&16 & 
1&00 & 
2&58 & 
0&07 & 
0&25 & 
1&24 \\
\oiii & 
5008 & 
0&44 & 
3&08 & 
8&02 & 
0&18 & 
0&76 & 
3&90 & 
0&41 & 
2&90 & 
7&72 & 
0&16 & 
0&73 & 
3&81 \\
\ni & 
5201 & 
0&06 & 
0&06 & 
0&05 & 
0&05 & 
0&03 & 
0&01 & 
0&05 & 
0&05 & 
0&05 & 
0&04 & 
0&02 & 
0&01 \\
\nii & 
5756 & 
0&01 & 
0&01 & 
0&01 & 
0&01 & 
0&01 & 
0&00 & 
0&01 & 
0&01 & 
0&01 & 
0&01 & 
0&00 & 
0&00 \\
\hei & 
5877 & 
0&09 & 
0&08 & 
0&06 & 
0&10 & 
0&12 & 
0&10 & 
0&06 & 
0&06 & 
0&05 & 
0&06 & 
0&09 & 
0&09 \\
\nai & 
5898 & 
0&05 & 
0&08 & 
0&08 & 
0&02 & 
0&01 & 
0&00 & 
0&03 & 
0&05 & 
0&07 & 
0&01 & 
0&01 & 
0&00 \\
\oi & 
6302 & 
0&16 & 
0&25 & 
0&28 & 
0&09 & 
0&10 & 
0&07 & 
0&10 & 
0&16 & 
0&22 & 
0&05 & 
0&08 & 
0&06 \\
\oi & 
6366 & 
0&04 & 
0&06 & 
0&08 & 
0&02 & 
0&02 & 
0&02 & 
0&02 & 
0&04 & 
0&06 & 
0&01 & 
0&02 & 
0&02 \\
\nii & 
6550 & 
1&07 & 
1&25 & 
1&08 & 
0&75 & 
0&42 & 
0&13 & 
0&61 & 
0&77 & 
0&80 & 
0&43 & 
0&30 & 
0&11 \\
\hi & 
6565 & 
5&06 & 
4&65 & 
3&86 & 
5&01 & 
3&93 & 
3&39 & 
2&86 & 
2&86 & 
2&86 & 
2&86 & 
2&86 & 
2&86 \\
\nii & 
6585 & 
3&08 & 
3&45 & 
2&69 & 
2&21 & 
1&23 & 
0&24 & 
1&73 & 
2&11 & 
1&99 & 
1&26 & 
0&89 & 
0&20 \\
\hei & 
6680 & 
0&02 & 
0&02 & 
0&02 & 
0&02 & 
0&03 & 
0&03 & 
0&01 & 
0&01 & 
0&01 & 
0&01 & 
0&02 & 
0&02 \\
\sii & 
6718 & 
1&01 & 
1&12 & 
0&86 & 
0&78 & 
0&73 & 
0&38 & 
0&55 & 
0&66 & 
0&62 & 
0&43 & 
0&52 & 
0&32 \\
\sii & 
6733 & 
0&78 & 
0&88 & 
0&69 & 
0&58 & 
0&52 & 
0&26 & 
0&42 & 
0&52 & 
0&50 & 
0&32 & 
0&37 & 
0&22 \\
\ariii & 
7138 & 
0&06 & 
0&11 & 
0&15 & 
0&04 & 
0&06 & 
0&07 & 
0&03 & 
0&06 & 
0&10 & 
0&02 & 
0&04 & 
0&06 \\
\oii & 
7322 & 
0&06 & 
0&06 & 
0&06 & 
0&05 & 
0&04 & 
0&03 & 
0&03 & 
0&03 & 
0&04 & 
0&02 & 
0&03 & 
0&03 \\
\oii & 
7332 & 
0&03 & 
0&04 & 
0&05 & 
0&02 & 
0&03 & 
0&02 & 
0&01 & 
0&02 & 
0&03 & 
0&01 & 
0&02 & 
0&02 \\
    \hline
  \end{tabular}
\end{table*}

Compared to the results that can be obtained by simply co-adding some
modest number of observed spectra, these reconstructed spectra are
purer tracers of either just AGN-like properties or just
star-forming-region properties, with a higher S/N and better
subtraction of the background galaxies. The table illustrates the wide
range of emission lines that can be measured reliably.  In the
following subsections we investigate the degree to which physical
information is carried by the progression of emission-line properties
along these loci. We note that, although table~\ref{t:lineratio}
includes emission lines as faint as 1 per cent of the H$\beta$ flux, the
following investigations rely only on the stronger lines.

\subsubsection{The AGN locus}

We compare the ICA reconstructions to a series of composite models of
extended emission regions photoionized by a central continuum
source. We follow the Locally Optimally-emitting Cloud (LOC) approach
used by \citet[][hereafter F97]{Ferguson97} in a similar study, in
which the narrow line region (NLR) is modelled as a large number of
individual clouds having power law distributions of distance, $r$,
from a central continuum source and of gas density, $n$. Here we
present a brief summary of our results to date; the full study will be
described in a future paper.

As was done in F97 (see also \citealt{Baldwin95}), we model the
integrated emitted spectrum of a large number of ionized clouds
distributed over a wide range in $\log(r)$ and $\log(n)$ with radial
distance and density distributions, $f(r) \propto r^\gamma$ and $g(n)
\propto n^\beta$, respectively.  We use version 10.00 of the plasma
simulation code {\sc Cloudy} \citep{Ferland98} to compute the
properties of the individual clouds. Our work builds on the F97
results by optimizing the continuum SED and the chemical abundances in
the gas to fit the ICA reconstruction corresponding to a2, the most
extreme AGN case.

\begin{figure*}
\includegraphics[width=\xsizedouble]{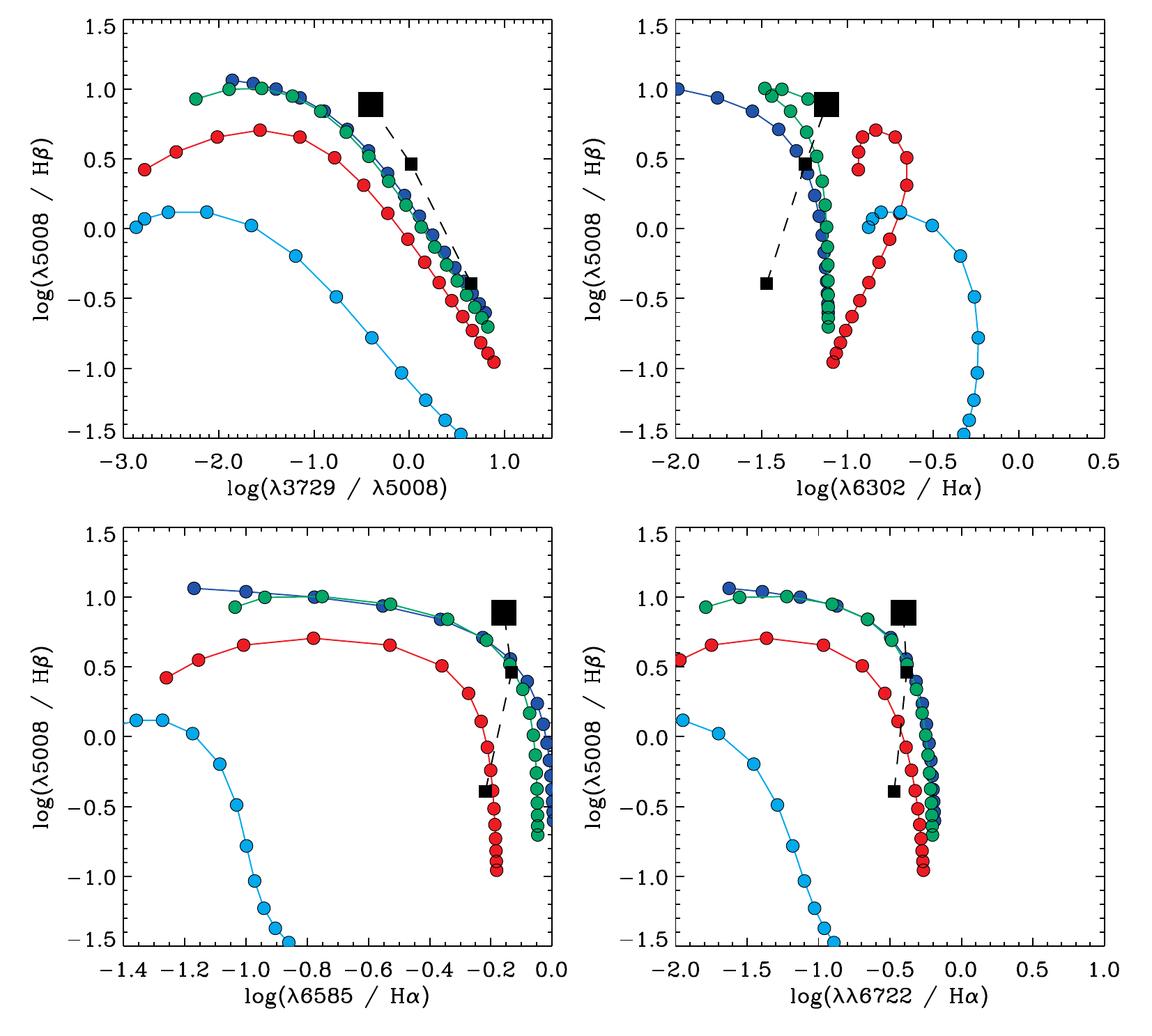}
 \caption{A subset of line-ratio diagrams are given here in the spirit
   of F97. In each panel the filled squares connected by the dashed
   line are the dereddened measurements from the a2--a1--a0 sequence
   of reconstructed spectra, with the large square representing the a2
   point. The free parameters are displayed for $\beta = -0.6$ (aqua),
   $-1.0$ (red), $-1.4$ (green), $-1.8$ (blue) and for $-2.0 \leq
   \gamma \leq 2.0$ in increments of 0.25 where the most negative
   $\gamma$ values correspond to the highest values along the
   $y$-axis. The a2 points are successfully fitted by $(\beta = -1.4,
   \gamma = -0.75)$, and the a1 points by $(\beta = -1.4, \gamma =
   -0.25)$, but the physical properties of the a0 reconstruction remain
   unclear.}
 \label{f:agn_models}
\end{figure*}

Fig.~\ref{f:agn_models} shows a subset of diagnostic emission line
intensity ratio diagrams calculated for our grids of models, which are
shown as filled circles connected by lines. The diagrams also show the
points measured from the reconstructed AGN locus, with the point
corresponding to the extreme AGN (a2) case shown as the larger filled
square. The LOC models shown here use an SED that has been adjusted to
improve the fit to the $I({\rm \hei}\,\lambda4687)/I({\rm H}\beta)$
ratio, which is a well-known SED indicator \citep{FO86}.  An ionizing
luminosity, $L_{\rm ion}=10^{43.5}$\,erg\,s$^{-1}$, UV temperature
cutoff of $T_{\rm cut}=2.5\times10^5$\,K, and X-ray to UV ratio of
$\alpha_{\rm OX}=-1.4$ provide a high-energy edge of the Big Blue Bump
that occurs at a somewhat lower energy than in the usual \citet{MF87}
AGN continuum; the sensitivity of different AGN line ratios to the SED
will be discussed in detail in our forthcoming paper. These LOC models
use roughly solar abundances (from F97) except that the N/H
abundance ratio has been increased to improve the fit to the $I({\rm
\nii}\,\lambda6585)/I({\rm H}\alpha)$ ratio. The final abundances
relative to hydrogen are: ${\rm He}=-0.987$, ${\rm Li}=-8.69$, ${\rm
Be}=-10.58$, ${\rm B}=-9.12$, ${\rm C}=-3.61$, ${\rm N}=-3.73$, ${\rm
O}=-3.31$, ${\rm F}=-7.52$, ${\rm Ne}=-3.92$, ${\rm Na}=-5.68$, ${\rm
Mg}=-4.42$, ${\rm Al}=-5.51$, ${\rm Si}= -4.49$, ${\rm P}=-6.43$,
${\rm S}=-4.80$, ${\rm Cl}=-6.72$, ${\rm Ar}=-5.60$, ${\rm K}=-6.87$,
${\rm Ca}= -5.65$, ${\rm Sc}=-8.80$, ${\rm Ti}=-6.96$, ${\rm
V}=-7.98$, ${\rm Cr}=-6.32$, ${\rm Fe}=-4.54$, ${\rm Co}= -7.08$,
${\rm Ni}=-5.75$, ${\rm Cu}=-7.73$, ${\rm Zn}=-7.34$.

The fits to the a2 point are not perfect, but for $\beta = -1.4$ and
$\gamma = -0.75$ are correct to within a factor of two. The fits to the
points further down the AGN locus remain unclear, although for a1
values of $\beta = -1.4$ and $\gamma = -0.25$ appear to give the best
fit.  The a0 reconstruction does not fit neatly on these sequences
and is likely to represent a composite of more than one source of
excitation.  The manner in which the loci were constructed limits the
potential contribution from star formation, but contributions due to
shock-heated gas or excitation due to LINERs remain possibilities.  In
contrast, the reconstructed MFICA spectra along the upper
part of the AGN sequence can be understood physically in terms of a
rather simple NLR model with photoionization by a single central
engine.

\subsubsection{The star formation locus}

Table~\ref{t:lineratio} and Fig.~\ref{f:templates} show that the
\hei\,$\lambda5877/$H$\beta$ ratio is in the range 0.09 for the s2 and
s1 reconstructions to 0.06 for the s0 reconstruction.  Here we show
that this ratio can be used to estimate the starburst age.

In \hii\ regions optical \hei\ and H$\beta$ lines form by
recombination from He$^+$ and H$^+$. The \hei$/$H$\beta$ intensity
ratio is proportional to the ratio of abundances of these ions. The
small \hei$/$H$\beta$ ratios found here are unlikely to reflect a
truly low He$/$H abundance ratio. The lower bound to the range of
He$/$H that can occur in a galaxy is that produced by Big Bang
nucleosynthesis (\citealt{AGN2}, Chapter 9), ${\rm He}/{\rm H} \sim
0.08$. If both He and H are singly ionized this abundance ratio
corresponds to $I({\rm \hei}\,\lambda5877) / I({\rm H}\beta) \sim
0.12$, where the recombination coefficients listed in \citet{AGN2} are
adopted, and pure recombination is assumed.  There will be a
collisional contribution which will increase the ratio for some of the
denser models considered below.

The observed intensity ratio is smaller than this lower limit,
suggesting that the He$^+/$H$^+$ ratio is smaller than the He$/$H
abundance ratio. Helium has two ionized states, He$^+$ and He$^{++}$,
and high-ionization nebulae can have significant amounts of He$^{++}$,
which produces \heii\ emission. However the lack of
\heii\,$\lambda$4687 emission shows that this is not important in our
sample and is consistent with the emission coming from \hii\ regions
rather than AGN or evolved objects like planetary nebulae. This means
that helium must be present in either atomic or singly ionized form in
regions where hydrogen is ionized.

It is most likely that there are significant parts of the
\hii\ regions where hydrogen is ionized but He is atomic and produces
no recombination emission. This happens in \hii\ regions ionized by
relatively cool stars because of the higher ionization potential of
helium. In this case the \hei$/$\hi\ intensity ratio is mainly set by
the SED, as we show next.

\begin{figure}
\includegraphics[width=\xsizesingle]{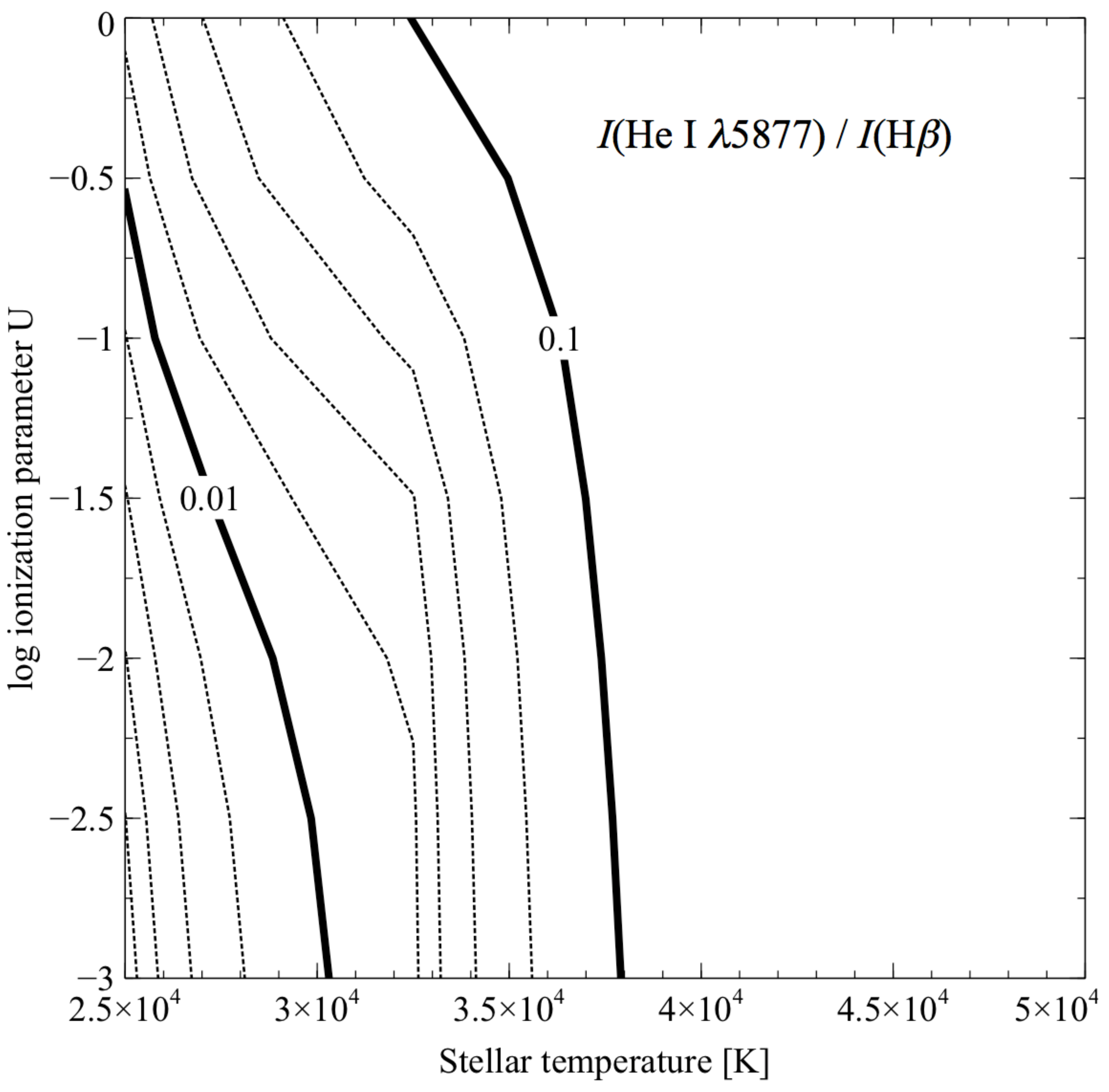}
 \caption{Intensity ratio $I({\rm \hei}\,\lambda 5877)/I({\rm
     H}\beta)$ for gas photoionized by stellar SEDs, for a range of
   ionization parameters and stellar effective temperatures.}
 \label{f:hiigrid}
\end{figure}

Fig.~\ref{f:hiigrid} quantifies the effect of the SED on the
\hei$/$\hi\ intensity ratio. In photoionization equilibrium the
ionization of the gas is determined both by the SED and by the
ionization parameter, $U$, the dimensionless ratio of ionizing photon
to hydrogen densities. The $x$-axis corresponds to varying SED shapes,
given as the stellar effective temperature assuming the TLUSTY grids
of atmospheres \citep{LH03,LH07}. The $y$-axis gives the log of the
ionization parameter, the second parameter that sets the ionization of
the gas. The contours show the \hei\,$\lambda5877/$H$\beta$ ratio
computed with {\sc Cloudy}. Galactic ISM abundances and dust were
assumed along with a hydrogen density of $10^3$\,cm$^{-3}$, typical of
\hii\ regions. The geometry was assumed to be a plane parallel layer,
a good approximation to blister \hii\ regions in 30 Dor
\citep*{PBF10}.

There are two regimes seen in Fig.~\ref{f:hiigrid}.  For stellar
temperatures higher than 38\,000\,K, corresponding to the O stars that
dominate the SED of the youngest star clusters, the intensity ratio
$I($\hei\,$\lambda5877)/I($H$\beta) \geq 0.12$ is nearly constant,
set by the He$/$H ratio in the host galaxy.  This regime fills the
right hand half of Fig.~\ref{f:hiigrid}.  For lower effective
temperatures, and hence older clusters, the situation becomes more
complex; in this regime the \hei$/$H$\beta$ ratio shows dependencies
on both the stellar temperature and the ionization parameter.

As the cluster ages and the SED grows softer it moves to the left in
Fig.~\ref{f:hiigrid}.  The \hei$/$H$\beta$ ratio will be constant
until the cluster reaches an age such that stars with $T_{\rm eff}
\geq 38\,000$\,K die, at which point the \hei$/$H$\beta$ ratio will
begin to decrease.  The hot-star calibration given by \citet*{HLH06},
which is also based on the TLUSTY stellar atmospheres used to make
Fig.~\ref{f:hiigrid}, shows that 38\,000\,K corresponds to an O6.5~V
star with a mass of 29\,M$_\odot$ and a main-sequence lifetime of
6\,Myr \citep{Schaller92}.

We further quantify the transition between the two regimes by
computing the SED produced by an evolving star cluster using a series
of Starburst99 \citep{Leitherer99,VL05} models.  Instantaneous
formation was assumed, using a \citet{Kroupa01} IMF truncated at 0.1
and 100\,M$_\odot$, and solar metallicity.  This fully specifies the
SED as a function of time.  In terms of fundamental parameters, the
ionization parameter -- which depends on the cluster luminosity, the
separation between the cluster and the gas, and the gas density -- is
the remaining unknown.  A real environment is likely to be highly
chaotic, much like the Magellanic Clouds, with shock-heated hot gas
intermixed with molecular clouds (\citealp{PBF10}; \citealp*{PBF11}).
Such an environment would not have a single ionization parameter, but
rather would be a mix of clouds with a wide range \citep{PBF11}, as is
described by LOC models.

\begin{figure}
  \includegraphics[width=\xsizesingle]{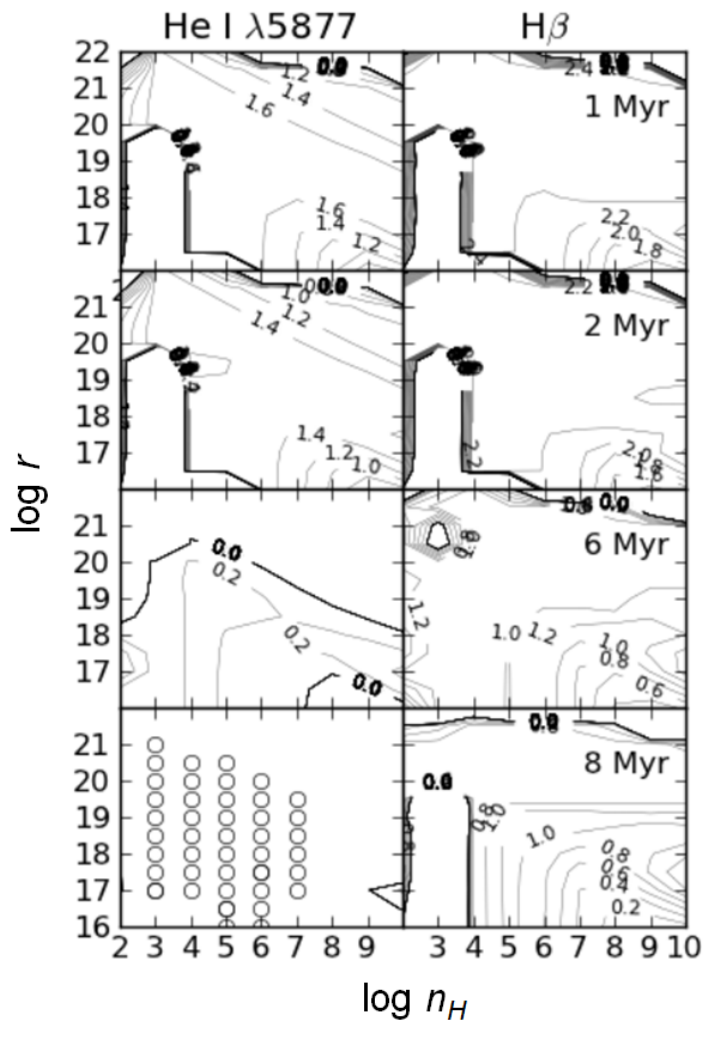}
  \caption{Contour plots showing the equivalent widths of
    \hei\,$\lambda5877$ and H$\beta$ lines emitted by gas clouds as a
    function of their hydrogen density $n_{\rm H}$ and their distance
    $R$ from starbursts of equal ionizing luminosity. The starbursts
    have different ages, as indicated in the right-hand panel in each
    row. The circles in the bottom-left panel show the $n_{\rm H}$,
    $R$ values of the curves plotted in Fig.~\ref{f:heihbage}.}
  \label{f:loccontours}
\end{figure}

Fig.~\ref{f:loccontours} shows contour plots of the
\hei\,$\lambda5877$ and H$\beta$ equivalent widths for starburst
models all having the same ionizing luminosity but with ages between 1
and 8\,Myr. The horizontal axis is the hydrogen density $n_{\rm H}$,
while the vertical axis is the distance $R$ from the ionizing
continuum source. Note the wide range of densities in the figure.  The
\hei\ lines will be enhanced by a collisional contribution for such
conditions \citep{Ferland86}. Lines of constant ionization parameter
$U \propto R^{-2} n_H^{-1}$ would run diagonally across these plots,
so the starburst age, ionization parameter space discussed above is
represented on Fig.~\ref{f:loccontours}. It is seen that while
H$\beta$ comes from most of the $n_{\rm H}$, $R$ plane for all ages,
the \hei\,$\lambda5877$ emission comes from only a restricted area by
the time an age of 6\,Myr is reached, and then is entirely absent by
8\,Myr.

\begin{figure}
  \includegraphics[width=\xsizesingle]{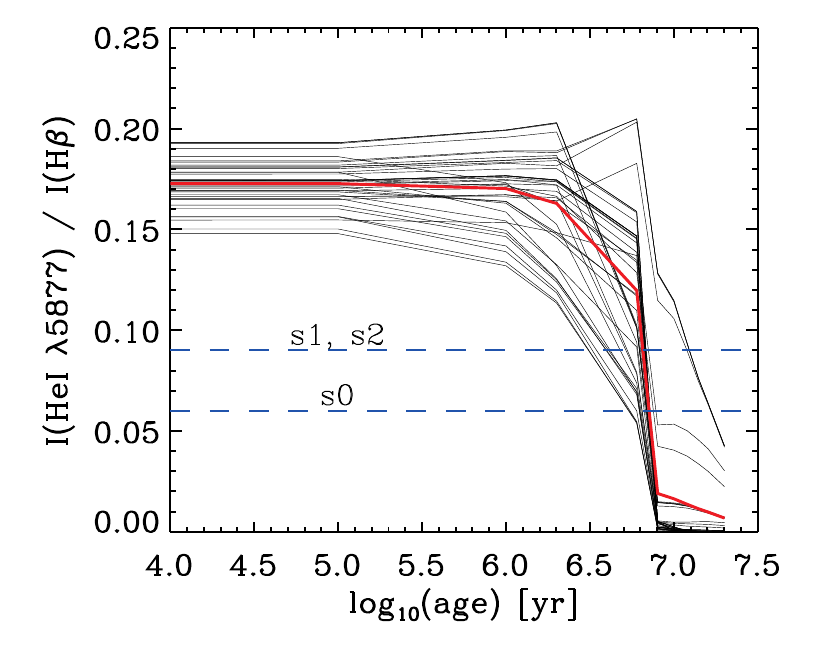}
  \caption{The \hei$/$H$\beta$ intensity ratio as a function of
    starburst age. The thin black lines are for each of the $n_{\rm
      H}$, $R$ points marked by the circles on the bottom-left panel
    in Fig.~\ref{f:loccontours}. The heavy red line is the average of
    all the thin lines. The horizontal dashed lines show the measured
    \hei$/$H$\beta$ ratios for ICA reconstructions s0, s1 and s2.}
  \label{f:heihbage}
\end{figure}

Fig.~\ref{f:heihbage} shows a series of plots of the \hei$/$H$\beta$
intensity ratio vs.\ age, made for the locations on the $n_{\rm H}$,
$R$ plane marked by the open circles in the bottom-left panel of
Fig.~\ref{f:loccontours}. The heavy red line on Fig.~\ref{f:heihbage}
is the average, corresponding to a LOC model with equal numbers of
clouds as a function of $\log(n_{\rm H})$ and $\log(R)$. The
\hei$/$H$\beta$ ratio integrated over the full $n_{\rm H}$, $R$ plane
will drop off sharply, as discussed above, with the exact cutoff age
depending on the distribution of the emitting clouds in $n_{\rm H}$
and $R$. The values \hei$/$H$\beta = 0.06$--$0.09$ measured for MFICA
reconstructions are consistent with a single starburst with an age of
about 7\,Myr. But they might also be due to a mix of star-forming
regions within each individual galaxy, with half of the H$\beta$
emission coming from gas ionized by stars sufficiently younger than
7\,Myr that their \hei$/$H$\beta$ ratio is at the maximum value, and
the other half coming from regions ionized by older stars with no
He$^+$ being formed. In a future paper we will test this hypothesis by
comparing detailed LOC models to the intensity ratios of the wide
variety of weak emission lines that can be measured using the MFICA
technique.

\section{Conclusions}

\label{s:conclusions}

In this paper we have presented an analysis of narrow emission-line
galaxies based on the MFICA technique.  A set of five continuum
components was generated from a mixed sample of galaxies with and
without emission lines.  Three of these components can be identified
with old, intermediate and young stellar populations, while the final
two components allow the reconstruction of the full range of ages of
stellar populations.  Using these components to fit and subtract the
continuum from a sample of emission-line galaxies, we then generated a
set of five emission-line components.  In combination, the five
continuum and five emission-line components can be used to produce
accurate reconstructions of the spectra of galaxies with a wide range
of properties.

We have provided a brief demonstration of the strong correlations
between the MFICA continuum weights and the VESPA star formation
histories presented by \citet{Tojeiro09}.  These correlations imply
that it will be possible to derive estimated star formation histories
for individual galaxies based on the MFICA results.

After identifying the regions of parameter space that correspond to
pure star formation and pure AGN, we made use of the high S/N MFICA
reconstructions to probe the physical conditions within these systems.
The most extreme AGN case is well fit by a model consisting of a large
number of ionized clouds with radial distance and density
distributions, $f(r)\propto r^\gamma$ and $g(n)\propto n^\beta$,
respectively, with $\gamma=-0.75$ and $\beta=-1.4$.  In the star
formation reconstructions, the measured \hei\,$\lambda5877/$H$\beta$
ratios imply a starburst age of about 7\,Myr, or a mix of star-forming
regions both older and younger than 7\,Myr.

A full analysis of the reconstructed spectra is deferred to a
forthcoming paper, but the initial investigations presented here serve
to illustrate the power of MFICA to probe the physical conditions
present within large samples of galaxies.  Techniques of this sort
will prove invaluable in the analysis of current and future
large-scale spectroscopic surveys.

\section*{Acknowledgments}

We are grateful to Manda Banerji for helpful discussions.  We thank
the anonymous referee for a thoughtful and in-depth review. JTA
acknowledges the award of an STFC Ph.D.\ studentship and an ARC Super
Science Fellowship.  PCH acknowledges support from the STFC-funded
Galaxy Formation and Evolution programme at the Institute of
Astronomy.  JAB and CTR acknowledge support from NASA ADP grant
NNX10AD05G and NSF grant AST-1006593.  GJF acknowledges support by NSF
(0908877, 1108928 and 1109061), NASA (07-ATFP07-0124, 10-ATP10-0053
and 10-ADAP10-0073), JPL (RSA No 1430426), and STScI (HST-AR-12125.01
and HST-GO-12309).  We wish to acknowledge the
support of the Michigan State University High Performance Computing
Center and the Institute for Cyber Enabled Research.

The Max Plank institute for Astrophysics/John Hopkins University
(MPA/JHU) SDSS data base was produced by a collaboration of researchers
(currently or formerly) from the MPA and the JHU. The team is made up of
Stephane Charlot (IAP), Guinevere Kauffmann and Simon White (MPA),
Tim Heckman (JHU), Christy Tremonti (U.\ Wisconsin-Madison -- formerly
JHU) and Jarle Brinchmann (Leiden University -- formerly MPA).

Funding for the SDSS and SDSS-II has been provided by the Alfred
P. Sloan Foundation, the Participating Institutions, the National
Science Foundation, the U.S. Department of Energy, the National
Aeronautics and Space Administration, the Japanese Monbukagakusho, the
Max Planck Society, and the Higher Education Funding Council for
England. The SDSS Web Site is http://www.sdss.org/.

The SDSS is managed by the Astrophysical Research Consortium for the
Participating Institutions. The Participating Institutions are the
American Museum of Natural History, Astrophysical Institute Potsdam,
University of Basel, University of Cambridge, Case Western Reserve
University, University of Chicago, Drexel University, Fermilab, the
Institute for Advanced Study, the Japan Participation Group, Johns
Hopkins University, the Joint Institute for Nuclear Astrophysics, the
Kavli Institute for Particle Astrophysics and Cosmology, the Korean
Scientist Group, the Chinese Academy of Sciences (LAMOST), Los Alamos
National Laboratory, the Max-Planck-Institute for Astronomy (MPIA),
the Max-Planck-Institute for Astrophysics (MPA), New Mexico State
University, Ohio State University, University of Pittsburgh,
University of Portsmouth, Princeton University, the United States
Naval Observatory, and the University of Washington.

\bibliography{../aux/bibtrunc}{}
\bibliographystyle{mn2e}

\end{document}